\documentclass[twocolumn,showpacs,amsmath,amssymb,prb,superscriptaddress]{revtex4}

\usepackage[pdftex]{graphicx} 
\usepackage{epstopdf}
\usepackage{verbatim}

\newcommand{\bra}[1]{\ensuremath{\langle#1|}}
\newcommand{\ket}[1]{\ensuremath{|#1\rangle}}

\def\up{_\uparrow}
\def\dn{_\downarrow}

\def\be{\begin{equation}}
\def\ee{\end{equation}}
\def\bea{\begin{eqnarray}}
\def\eea{\end{eqnarray}}

\usepackage{color}

\pacs{
73.21.Hb, 
71.10.Pm, 
74.78.Fk   
}

\begin{document}

\title{Interaction Effects in Topological Superconducting Wires Supporting Majorana Fermions}

\author{E.\ M.\ Stoudenmire}
\affiliation{Department of Physics and Astronomy, University of California, Irvine, CA 92697}
\author{Jason Alicea}
\affiliation{Department of Physics and Astronomy, University of California, Irvine, CA 92697}
\author{Oleg A.\ Starykh}
\affiliation{Department of Physics and Astronomy, University of Utah, Salt Lake City, Utah 84112}
\author{Matthew P.\ A.\ Fisher}
\affiliation{Department of Physics, University of California, Santa Barbara, CA 93106}

\date{\today}
\begin{abstract}
  Among the broad spectrum of systems predicted to exhibit topological superconductivity and Majorana fermions, one-dimensional wires with strong spin-orbit coupling provide one of the most promising experimental candidates.  Here we investigate the fate of the topological superconducting phase in such wires when repulsive interactions are present.  Using a combination of Density Matrix Renormalization Group, bosonization, and Hartree-Fock techniques, we demonstrate that while interactions degrade the bulk gap---consistent with recent results of Gangadharaiah \emph{et al}.---they also greatly expand the parameter range over which the topological phase arises.  In particular, we show that with interactions this phase can be accessed over a broader chemical potential window, thereby leading to greater immunity against disorder-induced chemical potential fluctuations in the wire.  We also suggest that in certain wires strong interactions may allow Majorana fermions to be generated without requiring a magnetic field.
\end{abstract}

\maketitle

\section{Introduction}

The search for Majorana fermions in condensed matter systems is presently being pursued with great vigor and optimism,\cite{WilczekPerspective,FranzPerspective,SternPerspective,Service:2011} driven to a large extent by the surge in realistic proposals\cite{ReadGreen,SrRu,FuKane,MajoranaQSHedge,FujimotoColdAtoms,SatoFujimoto,Sau,PatrickProposal,Linder,Alicea,HoleDopedMajoranas,Ghosh,Qi,1DwiresLutchyn,1DwiresOreg,ColdAtomMajoranas,HalfMetalNaCoO,Murakawa:2011} for stabilizing topological phases supporting these exotic particles.  Among the numerous platforms proposed, one-dimensional wires with strong spin-orbit interactions have emerged as a top contender for hosting Majorana modes.\cite{1DwiresLutchyn,1DwiresOreg} These systems possess the virtue that one can generate a topological superconducting state in a relatively simple manner, using only an applied magnetic field and proximity-induced Cooper pairing (which has already been achieved\cite{nanowireExpt}).  
Despite the fact that exchange statistics is ill-defined in one dimension, forming wire networks allows one to harness the non-Abelian braiding statistics of Majorana fermions\cite{ReadGreen,Ivanov,AliceaBraiding} and hence their potential for quantum computation,\cite{Hassler,AliceaBraiding,SauWireNetwork,TopologicalQuantumBus,Flensberg} even in this setting.  Thus the discovery of Majorana fermions in one-dimensional wires could ultimately lead to profound technological innovations.  

Much of the prior work on topological superconductivity in one-dimensional wires has focused on clean systems in the non-interacting limit.  A thorough understanding of the conditions required to experimentally observe and manipulate Majorana modes in these systems will likely require detailed studies incorporating both disorder and Coulomb repulsion, which poses a challenging problem.  Several recent works have considered the effects of disorder on the topological phase in a non-interacting wire;\cite{Multichannel1,Multichannel2,Multichannel3,Disorder1,Disorder2,Disorder3,Disorder4,Disorder5,Disorder6} interactions, however, have received less attention, even in the clean limit.  
There is a very natural reason to expect that Coulomb effects may influence the topological phase and its associated Majorana modes in important quantitative and perhaps even qualitative ways.  Namely, in the absence of proximity-induced superconductivity, the one-dimensional wires of interest in fact constitute repulsively interacting \emph{Luttinger liquids} whose properties can depart dramatically from those of a non-interacting wire.\cite{Sun:2007,Gangadharaiah:2008}  The interplay between Coulomb interactions, magnetism, and proximity-induced pairing in such a Luttinger liquid is far from obvious \emph{a priori} and very important to clarify for experiment.  

Gangadharaiah \emph{et al}.\ recently made an important advance in this direction using a field-theoretic approach.\cite{Gangadharaiah:2011}  More precisely, these authors addressed interaction effects in the limit where the Zeeman field is sufficiently large that only one of the spin-orbit-split bands plays a role.  Here the problem maps onto `spinless' fermions and can be studied analytically using bosonization.  Repulsive interactions were found to suppress the pairing-induced bulk gap in the wire, even destroying the gap altogether beyond a critical interaction strength\cite{Gangadharaiah:2011}---a rather physical but unfortunate finding from an experimental viewpoint.  Very recent work by Sela \emph{et al}.\ reached similar conclusions in a model relevant for quantum spin Hall edges.\cite{Sela:2011}

Here we follow up on the study of Gangadharaiah \emph{et al}., incorporating Coulomb repulsion at both weak and strong Zeeman fields using a combination of Density Matrix Renormalization Group (DMRG), Hartree-Fock, and bosonization methods.  Our DMRG simulations indicate that while interactions indeed suppress the bulk gap throughout, they also lead to a second, experimentally beneficial effect---topological superconductivity and Majorana fermions can be accessed at weaker magnetic fields and over a broader chemical potential window compared to the non-interacting limit.  Although we do not explicitly include disorder here, this result strongly suggests that interactions lead to greater immunity against chemical potential fluctuations in the wire, which is highly desirable given that this type of disorder is likely to provide one of the main obstacles to observing Majorana fermions in this setting.  

With moderately strong interactions (compared to the Zeeman field and pairing energies), we show that Hartree-Fock theory in fact performs remarkably well, capturing the DMRG phase diagram semi-quantitatively.  At strong interactions where Hartree-Fock proves inadequate, we employ bosonization to address the fate of the topological phase in a limit complementary to Ref.\ \onlinecite{Gangadharaiah:2011} as we treat both pairing \emph{and} the Zeeman field as perturbations.  Here too we show that the topological phase expands in phase space due to interactions (provided they are not sufficiently strong that pairing becomes irrelevant), and demonstrate consistency with DMRG simulations in this limit.  Finally, we suggest that a wire with strong Rashba and Dresselhaus couplings may be driven into a topological phase by interactions, even \emph{without} an applied magnetic field.  By performing DMRG simulations on a time-reversal invariant toy model with strong interactions, we show that the formation of Majorana fermions is indeed possible at zero magnetic field.

In the following section we describe the model system studied throughout, and set the stage by reviewing the physics in the non-interacting limit.  Interaction effects are discussed in Sec.\ III.  In Sec.\ IIIA we introduce the diagnostics we use for identifying the topological phase in DMRG, and present the phase diagram determined at finite Zeeman and pairing fields.  Our Hartree-Fock and bosonization studies are described in Secs.\ IIIB and C, respectively.  Section IIID compares our bosonization predictions with DMRG simulations, while Sec.\ IIIE discusses the possibility of accessing topological superconductivity in a time-reversal invariant system.  Finally, we summarize our results and discuss future directions in Sec.\ IV.  

\section{Model System}

The system we consider is a one-dimensional semiconducting wire with Rashba spin-orbit coupling of strength $\alpha$ and an orthogonal Zeeman field $V_z$.
Electrons in the wire additionally inherit an $s$-wave pairing field $\Delta$ (which we take to be real throughout) via the proximity effect with a neighboring bulk superconductor.  Including a local repulsive interaction of strength $U$,
we employ the following minimal continuum Hamiltonian for the system:
\begin{align}
\mathcal{H} & = \mathcal{H}_0 + \mathcal{H}_{\text{I}} \label{eqn:continuum_H}
 \\
\mathcal{H}_0 & = \int_{x} \psi^\dagger \left(-\frac{1}{2m} \partial^2_x - \mu - i \alpha \partial_x \sigma^y + V_z \sigma^z\right)\, \psi  \nonumber \\
& \mbox{} + \Delta \int_x \left( \psi\up \psi\dn + h.c. \right)  \\
\mathcal{H}_{\text{I}} & =  U \int_x \rho\up \rho\dn.
\end{align}
Here $x$ is the coordinate along the wire, $m$ is the effective mass for the electrons, and $\sigma^{x,y,z}$ are Pauli matrices.  The operator $\psi_{s}^\dagger$ creates an electron with spin $s = \uparrow,\downarrow$, while $\rho_s = \psi_s^\dagger \psi_s$ is the density operator.  Throughout, spin indices are summed whenever suppressed.

\begin{figure}[tp]
\includegraphics[width=\columnwidth]{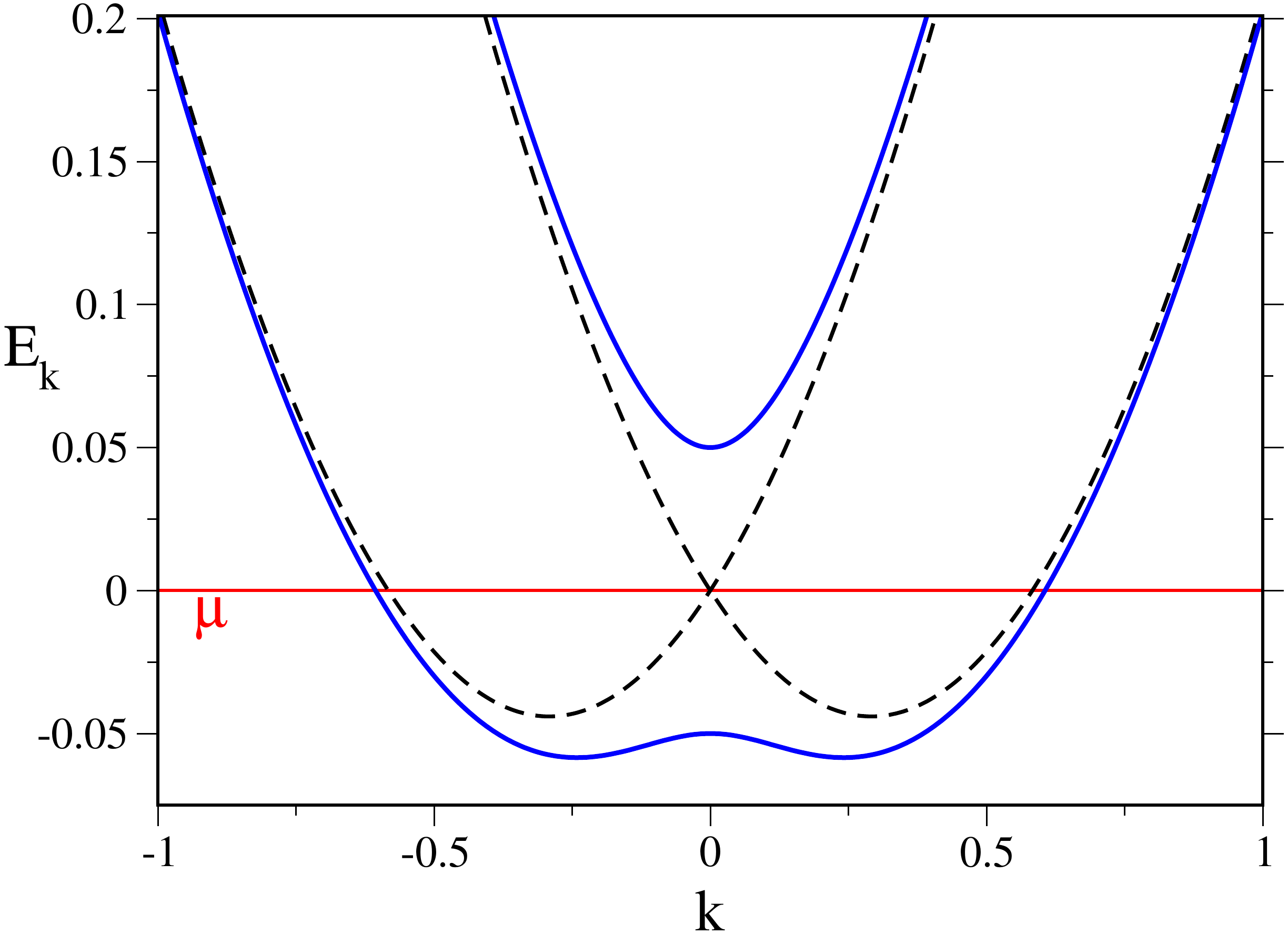}
\caption{Single-particle energy spectrum of the Hamiltonian $\mathcal{H}_0$ when $\Delta = 0$, $\alpha = 0.3$, and $m = 1$. The dashed lines show the spin-orbit-split bands in the time-reversal invariant limit with \mbox{$V_z=0$}.  Introducing a non-zero $V_z$ opens a gap in the spectrum at $k = 0$ as depicted by the solid lines.  This opens up a chemical potential window where the system exhibits only a single pair of Fermi points and thus resembles a `spinless' metal.  Turning on a weak $\Delta$ then effectively $p$-wave pairs the `spinless metal', generating a topological superconducting state supporting Majorana fermions. }
\label{fig:bands}
\end{figure}

Before turning to interaction effects, it will prove very useful to review the phase diagram of $\mathcal{H}$ in the non-interacting case.\cite{1DwiresLutchyn,1DwiresOreg}  Consider first the time-reversal-invariant limit with $V_z = \Delta = 0$.  Here $\alpha$ favors aligning the electron spins along or against the $y$-direction, depending on the momentum.  This produces the band structure shown by the dashed lines in Fig.\ \ref{fig:bands}, where the spins orient along the $+y$ and $-y$ directions
in the left and right parabolas, respectively.  The salient feature of these bands is the generic presence of four Fermi points in the spectrum, for arbitrary values of $\mu$ that are above the band minimum.  Turning on $\Delta \neq 0$ pairs $k$ and $-k$ states at both sets of Fermi points, producing an ordinary, non-topological superconductor.  The system always has a unique ground state here which is separated by a finite energy from all excited states.  
Note that as long as time-reversal symmetry remains unbroken, spin-orbit coupling plays no qualitative role, and the Hamiltonian can be smoothly deformed into that of a conventional 1D superconductor with $\alpha = 0$ (albeit with long-range order induced by the proximity effect).  

Restoring $V_z \neq 0$ breaks time-reversal symmetry and cants the spins along the $z$ direction.  This modifies the band structure in a crucial way; as shown by the solid lines in Fig.\ \ref{fig:bands}, a gap in the spectrum opens at $k = 0$.  In this case there exists a chemical potential window where the wire exhibits only \emph{two} Fermi points.  Here it is conceptually very useful to imagine projecting out the upper unoccupied band, and then viewing the wire as a `spinless' metal.  
Turning on a weak (compared to $V_z$) $s$-wave pair field $\Delta$ induces $p$-wave pairing between the fermions of the `spinless' metal, driving the system into a topologically non-trivial superconductor.  
In a finite system with open boundary conditions, exponentially localized Majorana modes $\gamma_1$ and $\gamma_2$ appear at the left and right ends of the wire.  Up to exponentially small corrections in their separation, $\gamma_1$ and $\gamma_2$ represent zero-energy modes and therefore lead to a two-fold ground state degeneracy, in contrast to the situation for the ordinary superconductor discussed above.  More precisely, if $|0\rangle$ is the ground state with even fermion parity, then $|1\rangle \propto \gamma_j|0\rangle$ is the ground state with odd parity.  All other states are separated by a finite energy gap from these two.  

This topological phase in fact connects smoothly to Kitaev's exactly solvable model for a $p$-wave superconducting chain, where the appearance of Majorana modes is very transparent.\cite{Kitaev:2001}  Note that spin-orbit coupling is absolutely crucial for the onset of this phase in the wire.  In particular, when $\alpha \neq 0$ the spins at the Fermi momenta $k_F$ and $-k_F$ in the lower band are \emph{not} parallel, which allows the spin-singlet pair field $\Delta$ to have a nontrivial effect on the `spinless' metal.  It will be useful to keep in mind, however, that as $V_z/\alpha$ increases the spins become increasingly aligned and hence $\Delta$ becomes correspondingly less effective at gapping the `spinless' metal.  

A quantitative analysis\cite{1DwiresLutchyn,1DwiresOreg} of the Bogoliubov-de Gennes equation with $U = 0$ reveals that the topological superconducting state exists only when the following criteria are satisfied:
\begin{equation}
\left\{
\begin{array}{c}   
    V_z > \Delta \\
    -\sqrt{V_z^2-\Delta^2}<\mu<\sqrt{V_z^2 -\Delta^2}
\end{array}\right.
\label{TopologicalCriteria}
\end{equation}
Outside of this regime, the system can not be viewed as a `spinless' metal either because the chemical potential lies outside of the Zeeman-induced gap or because $\Delta$ is sufficiently large that the upper and lower bands strongly couple to one another.  In either case the ordinary superconducting state emerges.  

One of the primary goals of this paper is understanding how this picture evolves when (possibly strong) interactions are present.  We will attack the problem with a combination of DMRG, Hartree-Fock, and bosonization methods.  To simulate the wire model using DMRG, we construct a lattice model that maps onto the continuum Hamiltonian above in the low-density limit.  This can be done in the usual way by writing $\mathcal{H}$ in momentum space and replacing
$k^2 \rightarrow 2(1-\cos k)$, $k\rightarrow \sin k$, $\int_k \rightarrow \frac{1}{L}\sum_k$, and $\psi_k \rightarrow \sqrt{L}c_k$ ($L$ is the system size).
In real space, the resulting lattice model reads
\begin{eqnarray}
  H &=& H_0 + H_I \label{H} \\
  H_0 &=& \sum_j \bigg{[} -\frac{t}{2} (c_j^\dagger c_{j+1} + h.c.) -(\mu-t)c_j^\dagger c_j \label{H0} \nonumber \\
  &-& \frac{\alpha}{2}(i c_j^\dagger \sigma^y c_{j+1} + h.c.) + V_z c_j^\dagger \sigma^z c_j 
  \nonumber \\
  &+& \Delta(c_{j\uparrow}c_{j\downarrow} + h.c.)\bigg{]}
  \\
  H_I &=& U\sum_j n_{j\uparrow} n_{j\downarrow},
\end{eqnarray}
where $n_{js} = c_{js}^\dagger c_{js}$.  Note that upon displaying $\hbar$ and the lattice constant $a$, the hopping strength is \mbox{$t = \hbar^2/(ma^2)$} while the continuum and lattice definitions of the spin-orbit coupling constant are related through \mbox{$\alpha_{\rm latt} = \frac{\hbar}{a} \alpha_{\rm cont}$}. After setting $\hbar = a = 1$ in what follows, $t = 1/m$ and $\alpha_{\rm latt} = \alpha_{\rm cont}$ so we employ the same symbol $\alpha$ to denote the spin-orbit strength in both models.  We will distinguish these for clarity only when discussing energy scales in the Hamiltonians momentarily.  When $U = 0$, the topological superconducting state again requires the criteria in Eq.\ (\ref{TopologicalCriteria}), exactly as in the continuum model.  

Since the Hamiltonians above contain many parameters, it is worth mentioning briefly the rough energy scales that are expected to be relevant in the problem.  Semiconductor effective masses are typically around $m \sim 0.05 m_e$, where $m_e$ is the bare electron mass.  Assuming the Rashba spin-orbit coupling in the wire is comparable to the values in two-dimensional quantum wells featuring heavy elements, we expect $\alpha_{\rm cont} \sim 10^4-10^5$m/s.  These estimates yield an energy $m \alpha_{\rm cont}^2 = \alpha_{\rm latt}^2/t \sim 1$K or so.  If a reasonable proximity effect can be established, then $\Delta$ can be on the scale of $1-10$K.  Fields of only a few Tesla can produce Zeeman energies up to $\sim 100$K in magnitude due to large spin-orbit enhancement of the $g$-factor.  And finally, $t$ determines the bandwidth and should therefore be on the eV scale.  These very rough estimates suggest that the relevant hierarchy of energies is $t \gg \alpha_{\rm latt} > \Delta$, with $V_z$ on the scale of $\alpha_{\rm latt}$ or smaller.  (Hereafter we cease to distinguish the lattice and continuum spin-orbit coupling constants.)  We will adhere to this rough guideline in our simulations, and henceforth work in units where $t = 1$.  Note, however, that as $\Delta$ and $V_z$ decrease toward zero, then finite-size effects caused by long coherence lengths make simulating the topological phase increasingly difficult.
\\

\section{Interaction Effects \label{sec:interaction_effects}}

\subsection{DMRG study of the phase diagram with interactions I \label{sec:DMRG_Study_I}}

As a first step towards incorporating interaction effects, we will use DMRG to assess the phase space occupied by the topological phase when $U > 0$.  More precisely, we would like to explore how the criteria displayed in Eq.\ (\ref{TopologicalCriteria}), which specify where the topological phase appears in the non-interacting limit, evolve with interactions.  Of particular experimental relevance is the required range of $\mu$, which is closely related to the stability of the topological phase against disorder-induced chemical potential variations which will inevitably be present in any semiconducting wire.  To protect against this kind of disorder, one would ideally  like to engineer the system to remain topological over as broad a range of $\mu$ as possible.  (There are, however, competing factors associated with the bulk gap discussed later in this section.)  Initially, we will address these issues when $\Delta$ is sufficiently large that the wire is rather far from the gapless Luttinger liquid regime.  In Secs.\ \ref{bosonization} and \ref{DMRGII} we will revisit the problem in the opposite regime when $\Delta, V_z \rightarrow 0$.  

All of our DMRG simulations are performed on a finite system with open boundary conditions.  The topological superconducting state can be identified numerically in three complementary ways.  First, since fermion parity is a good quantum number it is straightforward in DMRG to determine the minimum energies $E_{\rm even/odd}$ for eigenstates in the even and odd parity sectors.  The energy difference 
\begin{equation}
  \Delta E \equiv |E_{\rm odd} - E_{\rm even}|
\end{equation}
is finite in the ordinary superconducting state (because there is a unique ground state), but vanishes in the topological phase due to the presence of Majorana modes $\gamma_{1,2}$ exponentially localized at the left and right ends of the system.  Thus $\Delta E$ provides a sort of `order parameter' that distinguishes the topological and trivial superconducting phases.  

Second, because DMRG provides direct access to the many-body wavefunction we may compute the bulk entanglement spectrum. 
Defining the left reduced density matrix $\rho_L = \text{Tr}_R\ket{\Psi}\bra{\Psi}$ where the trace is over all sites in the right half of the wire,
the entanglement spectrum consists of the energies of the entanglement Hamiltonian $H_E = -\ln \rho_L$.
In the topological phase, this spectrum is expected to be two-fold degenerate.\cite{Turner:2010}
Physically, one may think of the entanglement degeneracy as a precursor of the zero-energy Majorana edge states that would appear 
if one were to cut the wire through a bulk bond.\cite{Qi:2011}

\begin{figure}[tp]
\includegraphics[width=\columnwidth]{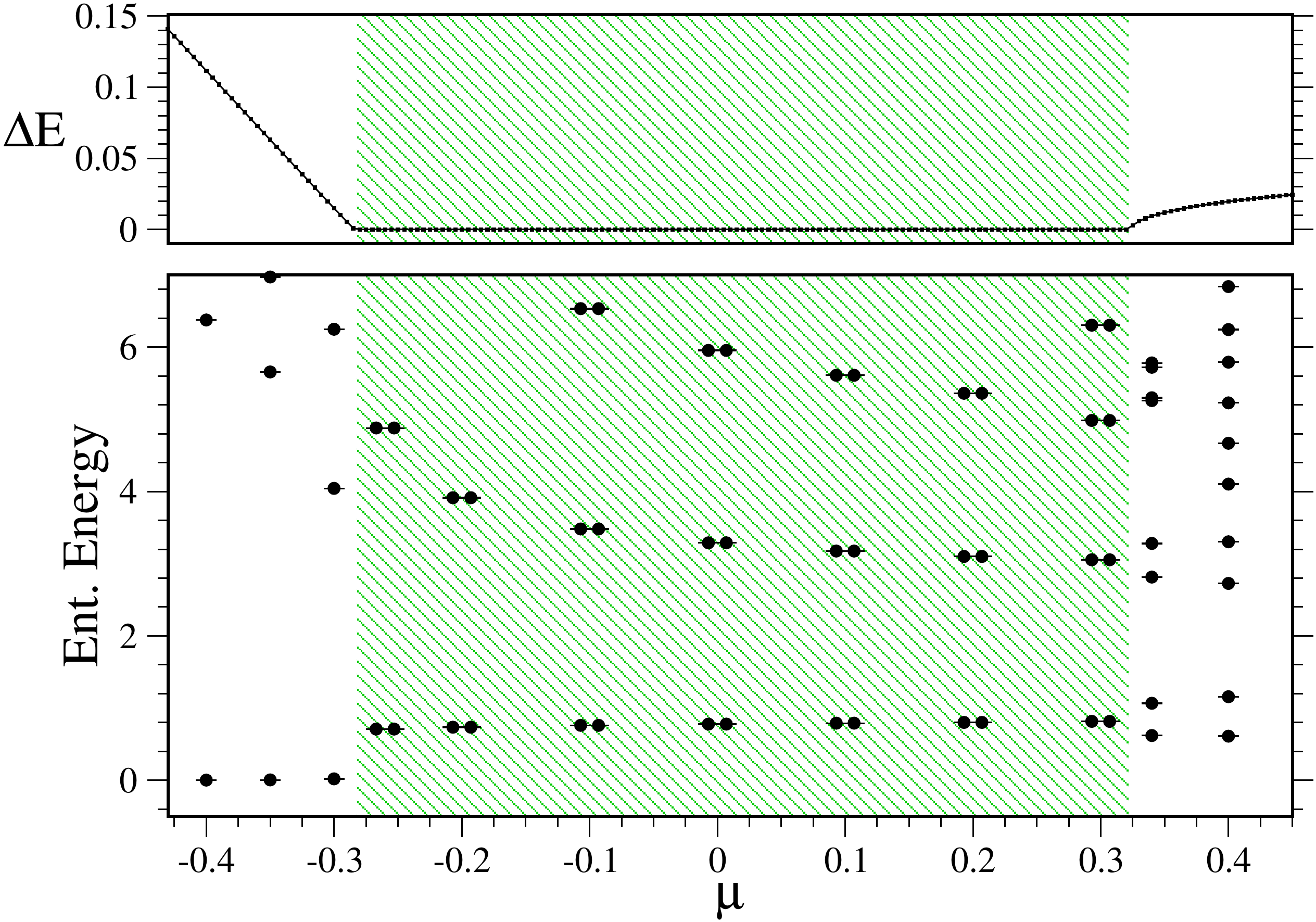}
\caption{Even/odd sector ground state energy difference $\Delta E$ (upper panel) and entanglement spectrum at the center bond (lower panel) for a 400 site chain with 
$\Delta=0.1$, $\alpha=0.3$, $V_z=0.3$ and $U=0.1$. The topological superconducting phase is signaled by \mbox{$\Delta E=0$} together with 
a two-fold degeneracy in the bulk entanglement spectrum.}
\label{fig:order_params}
\end{figure}

In Fig.~\ref{fig:order_params} we display $\Delta E$ (upper panel) and the entanglement spectrum at the center of a 
400 site chain (lower panel) as a function of $\mu$ with $\Delta=0.1$, $\alpha=0.3$, $V_z=0.3$
and $U=0.1$. The region of parameter space over which there is an entanglement degeneracy exactly corresponds to the topological phase with $\Delta E = 0$.

\begin{figure}[tp]
\includegraphics[width=\columnwidth]{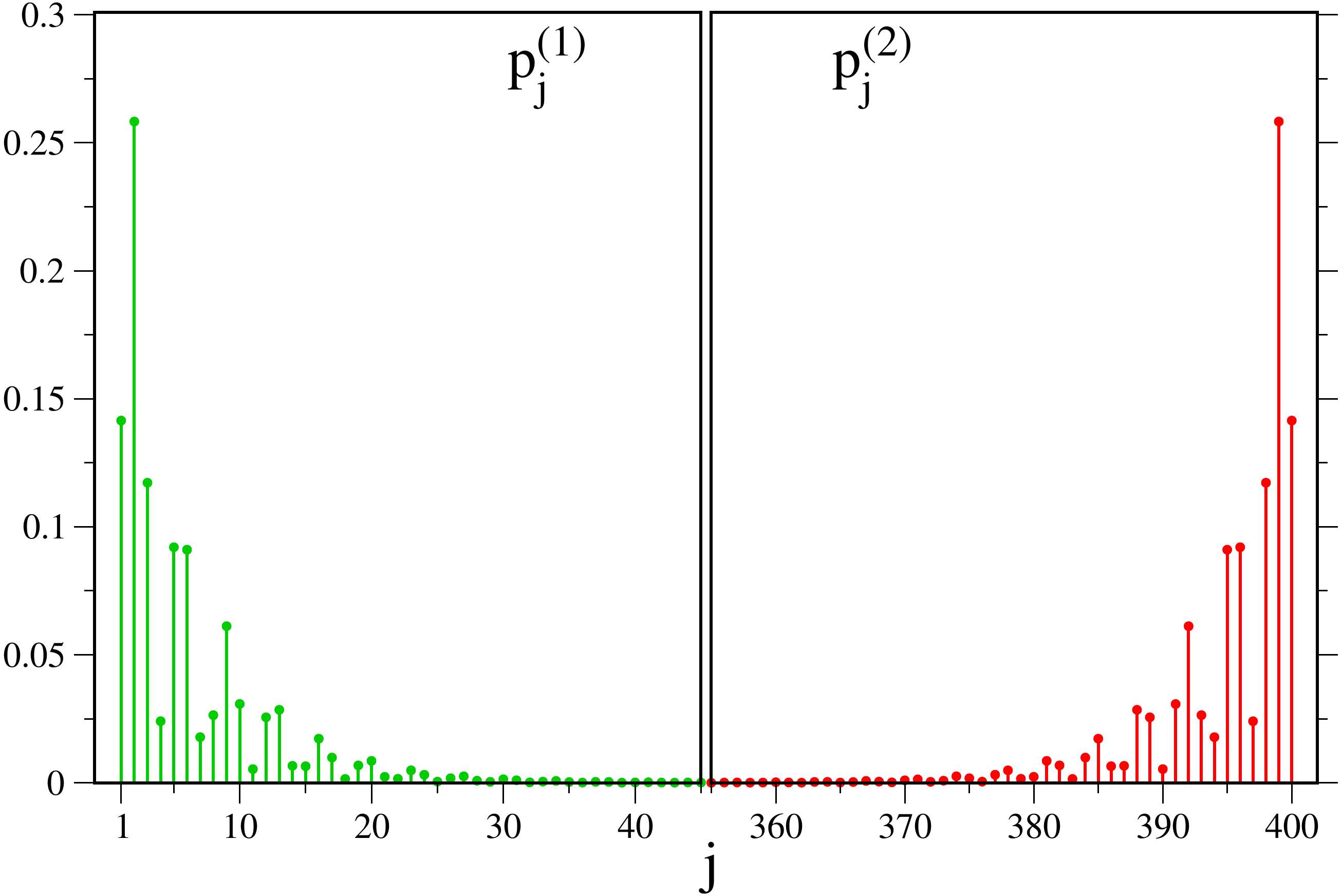}
\caption{Probability distribution $p^{(a)}_j = \sum_s |\phi_{js}^{(a)}|^2$ of the single-particle Majorana edge state wavefunctions $\phi^{(1)}$ (left panel) and $\phi^{(2)}$ (right panel)
for a 400 site chain with $\Delta=0.1$, $\alpha=0.3$, $V_z=0.3$, $U=0.1$ and $\mu=0$. For these parameter values the system is in the topological
phase (see Fig.~\ref{fig:order_params}), thus $\phi^{(1,2)}$ are non-zero only near a single edge of the system and decay exponentially into the bulk.}
\label{fig:majoranas}
\end{figure}

A third way to detect the topological phase numerically is by extracting the wavefunctions of the Majorana edge modes as follows.  Let $|0\rangle$ and $|1\rangle$ be the minimum-energy many-body wavefunctions for the wire in the even and odd parity sectors.  In the topological phase these are degenerate and can be related through the Majorana operators $\gamma_{1,2}$ that satisfy $\gamma_a^2 = 1$ and $\{\gamma_a,\gamma_{b}\} = 2\delta_{ab}$ according to
\begin{equation}
  |1\rangle = e^{i\theta}\gamma_1|0\rangle = i e^{i\theta}\gamma_2|0\rangle.
  \label{GroundStateRelations}
\end{equation}
Here $\theta$ is related to the overall phase factors for $|0\rangle$ and $|1\rangle$, and can always be absorbed into the definition of (say) $|0\rangle$.  Let us assume the following expansion for $\gamma_{1,2}$:
\begin{eqnarray}
  \gamma_a = \sum_j \sum_{s = \uparrow,\downarrow}(\phi_{js}^{(a)} c_{js} + h.c.),
  \label{MajoranaExpansion}
\end{eqnarray}
where $\phi^{(a)}$ is the wavefunction corresponding to Majorana mode $\gamma_a$.  The component $\phi^{(a)}_{is}$ is given by the anticommutator
\begin{equation}
  \{c_{is}^\dagger,\gamma_a\} = \phi^{(a)}_{is}.
  \label{anticommutator}
\end{equation}
Using Eqs.\ (\ref{GroundStateRelations}) and (\ref{anticommutator}), one can show that 
\begin{eqnarray}
  \phi_{is}^{(1)} &=& e^{i\theta}\langle 1 |c_{is}^\dagger|0\rangle + e^{-i\theta}\langle 0 |c_{is}^\dagger|1\rangle
  \\
  \phi_{is}^{(2)} &=& ie^{i\theta}\langle 1 |c_{is}^\dagger|0\rangle - ie^{-i\theta}\langle 0 |c_{is}^\dagger|1\rangle.
\end{eqnarray}
Thus measuring $\langle 1|c_{is}^\dagger |0\rangle$ and $\langle 0|c_{is}^\dagger |1\rangle$ numerically allows one to back out the precise form of the Majorana wavefunctions when one is in the topological phase. Figure~\ref{fig:majoranas} shows an explicit DMRG calculation of the probability distribution 
$p^{(a)}_j = \sum_s |\phi_{js}^{(a)}|^2$ of the Majorana wavefunctions, each of which are
indeed localized at a single edge in the topological phase.

A cautionary remark is in order.  The procedure outlined in the preceding paragraph relied on our assumption in Eq.\ (\ref{MajoranaExpansion}) that the Majorana operators are purely linear in the lattice fermion operators $c_{js}$ and $c_{js}^\dagger$.  This assumption of course holds in the non-interacting limit, though there is no obvious reason why higher-order terms (involving, \emph{e.g.}, three-fermion components) should be forbidden in an interacting system.  One can in fact write down simple exactly solvable interacting models where the Majorana operators involve \emph{only} three-fermion terms.  Nevertheless, in all our simulations, including those with very strong interactions, we empirically find that such corrections, if present at all in our model, are exceedingly weak.  This can be deduced by computing $\phi^{(1)}$ and $\phi^{(2)}$ as outlined above and then computing their normalization.  In all cases we examined the normalizations deviate from unity by less than 1\%, strongly suggesting that a single-body decomposition of the Majorana operators is indeed adequate.

\begin{figure}
\includegraphics[width=\columnwidth]{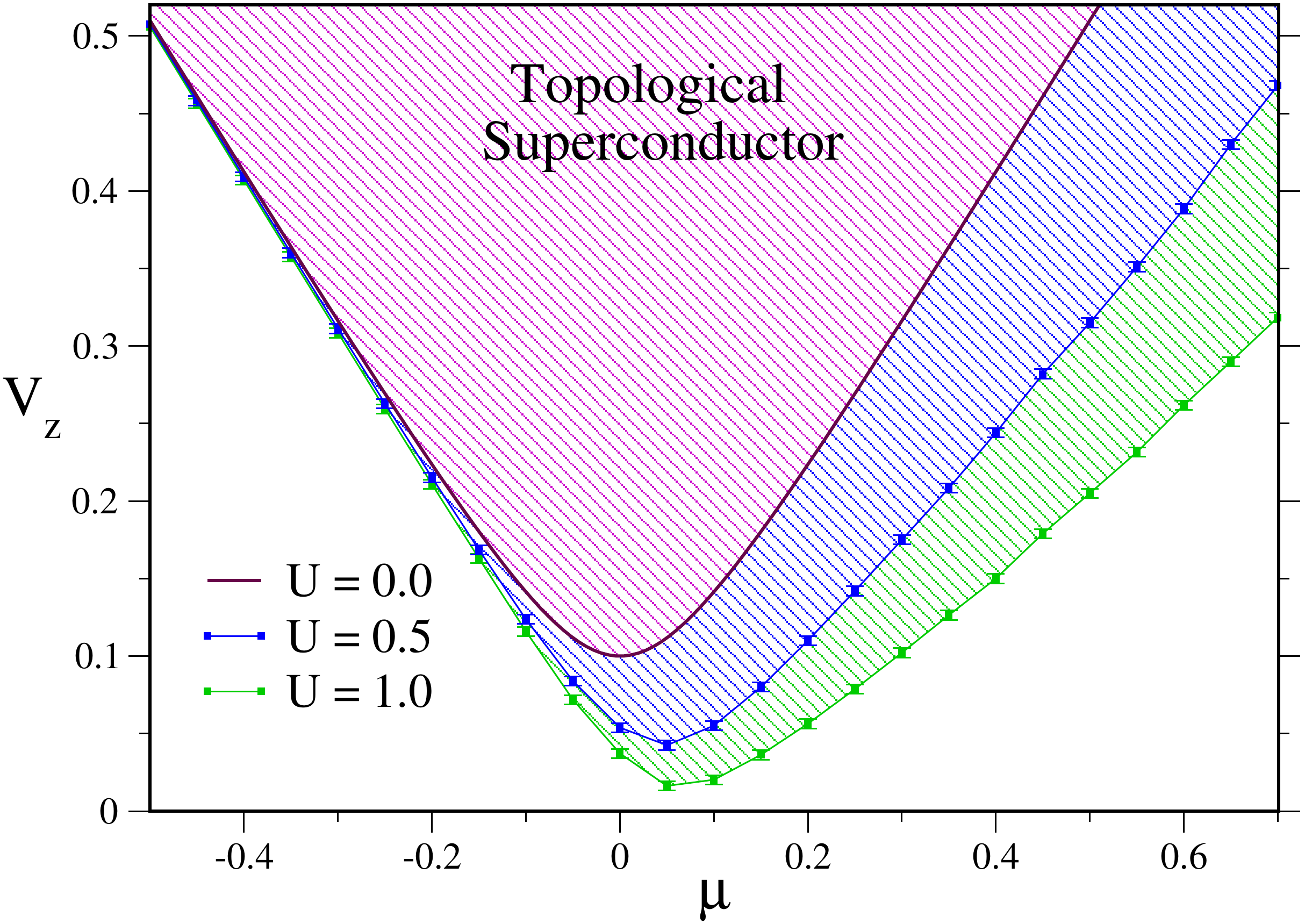}
\caption{Phase diagram showing the broadening of the topological superconductor phase (shaded regions) with increasing interaction strength. 
The data was obtained for a $400$ site chain with $\Delta=0.1$ and $\alpha=0.3$.}
\label{fig:phase_diagram}
\end{figure}

\begin{figure}
\includegraphics[width=\columnwidth]{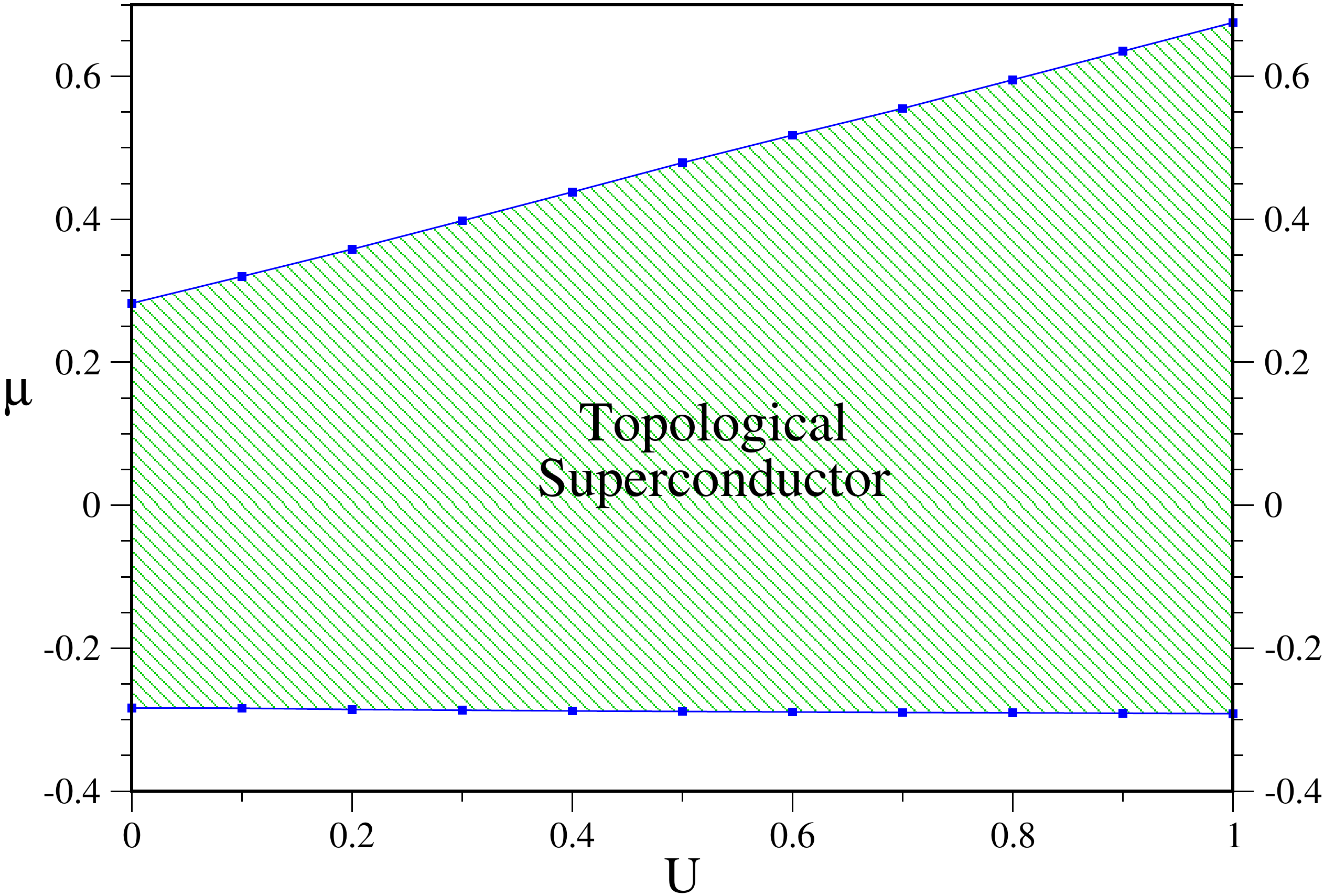}
\caption{Broadening of the topological superconductor phase at fixed Zeeman field of $V_z=0.3$ with increasing interaction strength. The data corresponds to  
a $400$ site chain with $\Delta=0.1$ and $\alpha=0.3$. Error bars are smaller than the symbol sizes.}
\label{fig:mu_range}
\end{figure}

Using these methods, one can employ DMRG to determine the parameter range in which the topological phase exists, for arbitrary-strength interactions.  Figure~\ref{fig:phase_diagram} illustrates the phase diagram for a 400-site chain with $\Delta = 0.1$ and $\alpha = 0.3$ as a function of $\mu$ and $V_z$, with three different interaction strengths.  The finite-$U$ phase boundaries were obtained by sweeping $V_z$ at fixed $\mu$, and plotting the value at which $\Delta E$ first vanishes.  Note that the left phase boundary changes very little with interactions since the electron density is extremely low in that region of the phase diagram.  The minimum of the topological phase boundary additionally shifts to finite $\mu$; this property arises because the $U$ repulsion adds a charging energy and thus effectively renormalizes the chemical potential.  Two more important trends are also evident in the figure: as $U$ increases, 1) the minimum value of $V_z$ required to stabilize the phase decreases significantly, and 2) at larger $V_z$ the topological phase occurs over a much broader chemical potential window.  The latter result is further illustrated in Fig.\ \ref{fig:mu_range}, which displays the phase boundary as a function of $U$ at fixed $V_z = 0.3$.  Both features are desirable from an experimental standpoint.  The decrease in the required $V_z$ in principle allows one to apply weaker fields, thereby disturbing the proximate superconductor less, while as already mentioned the widening of the topological phase as a function of $\mu$ implies greater immunity against chemical potential fluctuations in the wire.

The phase diagrams shown in Figs.\ \ref{fig:phase_diagram} and \ref{fig:mu_range} admit a natural interpretation
if we think of repulsive interactions as enhancing the effective Zeeman splitting while suppressing the pairing $\Delta$.
Electrons can avoid an energy cost from the repulsion $U$ by aligning their spins. Also, $U$ has the `wrong' sign for favoring pairing.  
According to Eqs.\ (\ref{TopologicalCriteria}), if introducing $U>0$ effectively enhances $V_z$ while suppressing $\Delta$, then qualitatively similar behavior to what we found numerically ought to occur.  However, this interpretation also suggests that the experimental benefits mentioned above are not without a cost.   
For example, the more spin-polarized the electrons are, the less effective the $s$-wave pair field $\Delta$ is at inducing pairing.  
Thus one expects a decrease in the bulk excitation gap in the topological phase.  

To test this picture we measure the effect of interactions on the bulk gap and magnetization within the topological phase.  Because of the ground state degeneracy, one must compute the lowest \emph{three} eigenstates of the interacting Hamiltonian to capture the bulk gap.  This poses a challenge for the usual DMRG technique of 
targeting multiple states as one would have to represent three states using the same basis. 
To overcome this difficulty, we instead compute the lowest three eigenstates one at a time, constraining each state to be orthogonal
to those previously calculated by introducing an energy penalty for any non-zero overlap. 
This requires us to be able to store and manipulate wavefunctions from independent DMRG calculations which we achieve by 
working in the matrix product state formalism.\cite{Schollwoeck:2011}

\begin{figure}[tp]
\includegraphics[width=\columnwidth]{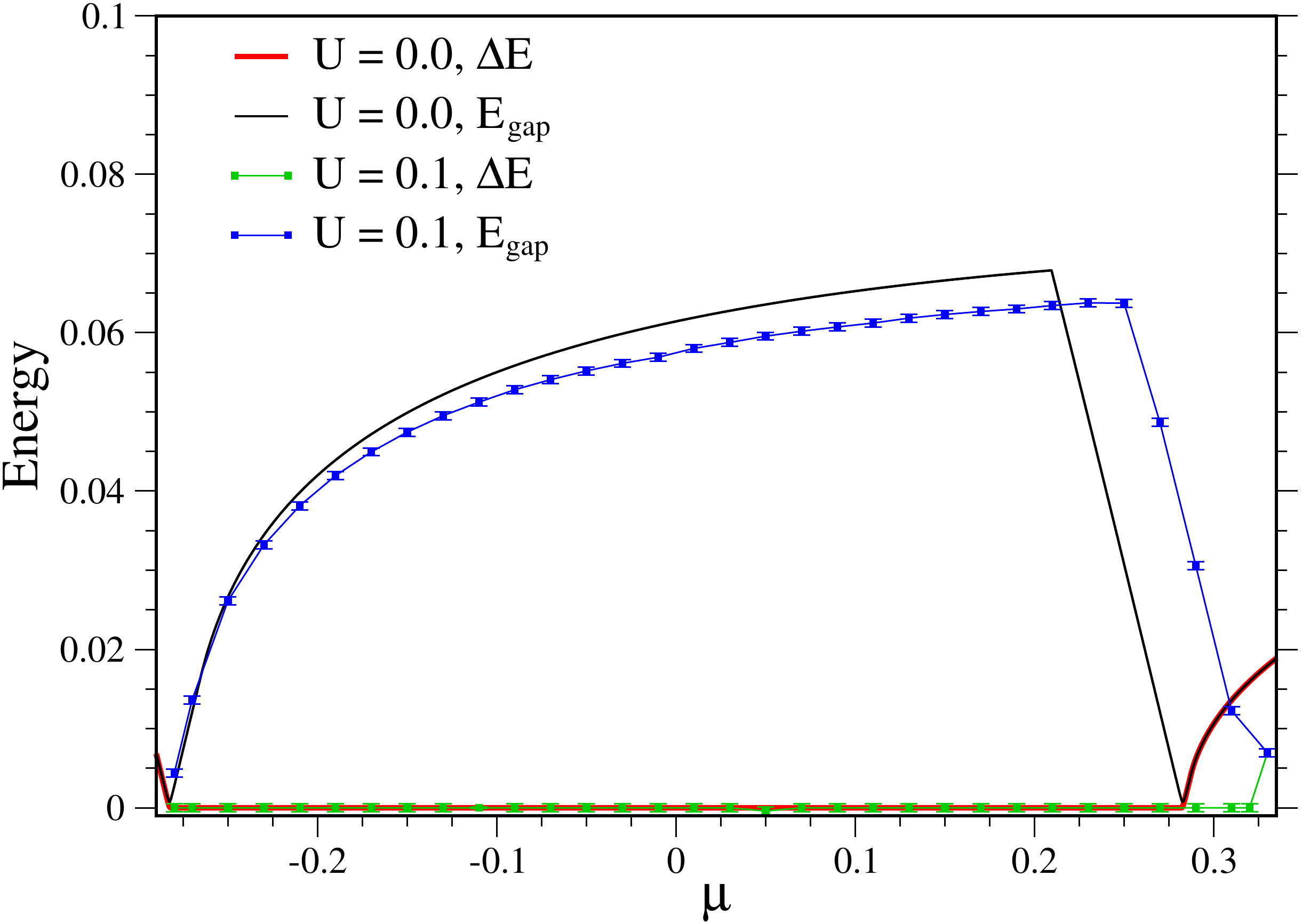}
\caption{Energy of the first and second excited states relative to the ground state for interaction strengths \mbox{$U=0$} and \mbox{$U=0.1$}.
In the topological phase the first excited state is degenerate with the ground state. The data was again obtained for a $400$ site chain with $\alpha=0.3$, $\Delta=0.1$ and $V_z=0.3$.}
\label{fig:gap}
\end{figure}

Figure~\ref{fig:gap} shows how $\Delta E$ and the bulk gap $E_{\rm gap}$ depend on $\mu$ for $U=0$ and $U=0.1$, keeping the other parameters fixed to $\Delta = 0.1$, $\alpha = 0.3$, and $V_z = 0.3$. In order to be sure that the gap we obtain is truly characteristic of the bulk, rather than arising from an edge excitation,
we have checked that local properties of the second excited state, such as the electron density, are significantly different from the ground state throughout the system.
The bulk gap turns out to be roughly constant across the topological phase, changing rapidly only near 
the phase boundaries.  (In an infinite system, the bulk gap can be set either by excitations carrying zero momentum or momentum near the Fermi points, depending on parameters.  This accounts for the abrupt change in slope near $\mu= 0.2$ in the figure.)  
We can therefore treat the maximum value of the gap as a measure for the robustness of the topological phase (to thermal fluctuations).

\begin{figure}[b]
\includegraphics[width=\columnwidth]{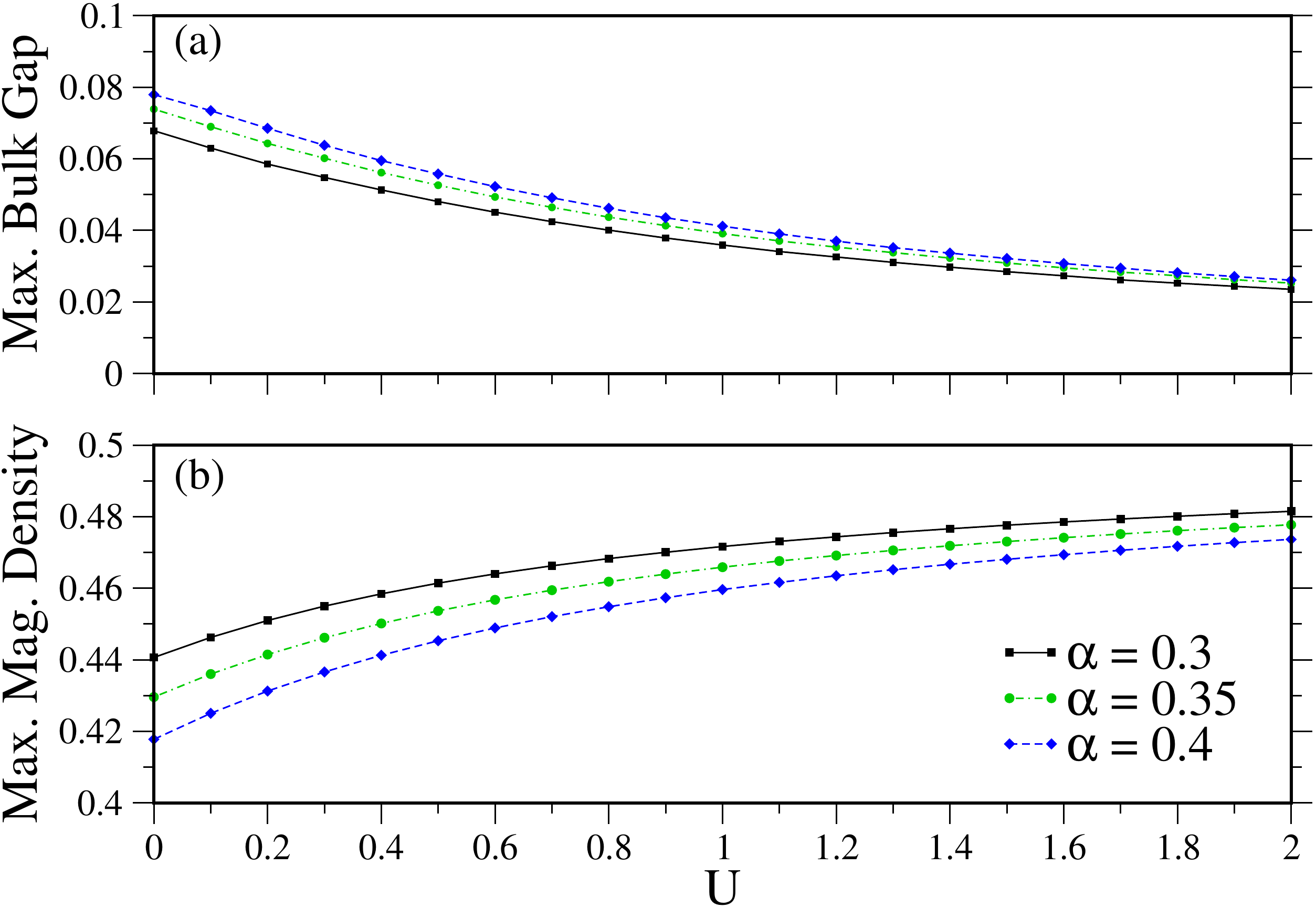}
\caption{Interaction dependence of (a) the maximum bulk energy gap and (b) the maximum magnetization per particle within the topological phase 
for a $400$ site chain with parameters $\Delta=0.1$ and $V_z=0.3$. Error bars are smaller than the symbol sizes.}
\label{fig:gap_magpp}
\end{figure}

Figure~\ref{fig:gap_magpp}(a) shows this maximum gap value as a function of interaction strength for various values of the spin-orbit
coupling. Consistent with the prediction of Gangadharaiah \emph{et al}.\cite{Gangadharaiah:2011} for the large-$V_z$ limit and the physical picture discussed above, we find that the bulk gap monotonically decreases as the interaction strength increases. The actual values observed range from about 20\% to 80\% of the pairing field $\Delta$.  

Next, to check the intuition that taking $U>0$ should lead to greater spin alignment, we measure the magnetization per particle which, like the bulk gap, is roughly constant as a function of $\mu$ over most of the topological phase.  We can therefore again take its maximum value as an indication of the magnetization for the phase.  The data appears in Fig.\ \ref{fig:gap_magpp}(b) and clearly indicates that interactions effectively enhance the Zeeman field.

\subsection{Hartree-Fock analysis}
\label{sec:HF}

Remarkably, qualitative and even semi-quantitative agreement with the phase diagram calculated from DMRG can be obtained within a simple Hartree-Fock approach.  Of course Hartree-Fock theory is not known for its spectacular success in one-dimensional systems, where fluctuations typically negate its predictions even on a qualitative level.  Here, however, the presence of spin-orbit coupling, the Zeeman field, and pairing conspire to eliminate all continuous symmetries for the problem, rendering the system rather `stiff' to fluctuations.  Consequently Hartree-Fock theory is expected to perform reasonably well here, at least when these symmetry-breaking fields are sufficiently strong.   

To facilitate comparison with the DMRG results above, we continue to work with a lattice model, though now with periodic boundary conditions.  We will consider a restricted set of Hartree-Fock trial states which are ground states of the following effective \emph{non-interacting} Hamiltonian,
\begin{eqnarray}
H_{\rm eff} &=& \sum_j \bigg{[}-\frac{\tilde t}{2} (c_j^\dagger c_{j+1} + h.c.) -(\tilde\mu-\tilde t)c_j^\dagger c_j 
  \nonumber \\
  &-& \frac{\tilde \alpha}{2}(i c_j^\dagger \sigma^y c_{j+1} + h.c.) + \tilde V_z c_j^\dagger \sigma^z c_j 
  \nonumber \\
  &+& \tilde \Delta(c_{j\uparrow}c_{j\downarrow} + h.c.)\bigg{]}.
\end{eqnarray}
Notice that this has the same form as $H_0$ in Eq.\ (\ref{H0}), except that now $\tilde t, \tilde \mu, \tilde \alpha, \tilde V_z$, and $\tilde \Delta$ are regarded as variational parameters.  One can diagonalize $H_{\rm eff}$ by going to momentum space and then expressing
\begin{eqnarray}
  c_{k \uparrow} &=& \varphi_1^A(k) d_{Ak} + \varphi_1^B(k) d_{Bk} + \varphi_3^A(k) d_{A-k}^\dagger + \varphi_3^B(k) d_{B-k}^\dagger
  \nonumber \\
  c_{k \downarrow} &=& \varphi_2^A(k) d_{Ak} + \varphi_2^B(k) d_{Bk} + \varphi_4^A(k) d_{A-k}^\dagger + \varphi_4^B(k) d_{B-k}^\dagger. \nonumber \\
\end{eqnarray}
Here $d_{A/Bk}^\dagger$ create quasiparticle excitations with energy $E_{A/B}(k)>0$ and $\phi^{A/B}_j(k)$ are the corresponding wavefunction components.  While the energies and wavefunctions can be obtained analytically from the Bogoliubov-de Gennes equation, here we simply note that due to spin-orbit coupling $\phi^A_{2,3}$ and $\phi^B_{2,3}$ are odd in $k$ while all other components are even.  The ground state of $H_{\rm eff}$ is annihilated by all $d_{A/B k}$ operators and can thus be written in terms of the vacuum $|{\rm vac}\rangle$ of $c_{js}$ fermions as
\begin{equation}
  |\psi\rangle = \prod_k d_{Ak}d_{Bk}|{\rm vac}\rangle.
\end{equation}
(This ground state is unique in the periodic boundary condition geometry.)  

The variational parameters are selected by minimizing the expectation value of the interacting Hamiltonian $H$ with respect to the trial state $|\psi\rangle$,
\begin{eqnarray}
  E_{HF} &=& E_0 + E_I 
  \\
  E_0 &=& \langle \psi|H_0|\psi\rangle
  \\
  E_I &=& \langle \psi|H_{I}|\psi\rangle.
\end{eqnarray}
The $H_0$ expectation value is of course minimized when the variational parameters equal their respective values in the original, interacting Hamiltonian.  Deviations away from these values may become favorable, however, when an increase in $E_0$ is offset by an energy gain from $E_I$.  The latter may be explicitly expressed as
\begin{eqnarray}
  E_I &=& \frac{U}{L}(N_{\uparrow} N_{\downarrow} + |A|^2),
  \label{EI}
\end{eqnarray}
where $N_s$ is the average fermion number with spin $s$ and $A$ is the anomalous correlator that is non-zero due to pairing:
\begin{eqnarray}
  N_{s} &=& \sum_{k} \langle c_{ks}^\dagger c_{ks}\rangle 
  \\
  A &=& \sum_{k} \langle c_{-k\downarrow} c_{k\uparrow}\rangle.
\end{eqnarray}
(The exchange term $-|\sum_{k}\langle c_{k\uparrow}^\dagger c_{k\downarrow}\rangle|^2$ vanishes by symmetry, which can be seen formally due to the parity of the wavefunction components mentioned above under $k\rightarrow -k$.)  Thus there are two ways in which the system may lower its interaction energy: by enhancing its spin polarization to decrease the first term in Eq.\ (\ref{EI}) and by pairing less strongly to diminish the second term in Eq.\ (\ref{EI}).  

We deduce the outcome of the competition between $E_0$ and $E_I$, and hence the fate of the topological phase, by numerically minimizing the Hartree-Fock energy $E_{HF}$ over a range of $V_z$ and $\mu$.  Since the overall scale of the parameters in $H_{\rm eff}$ does not affect $E_{HF}$, we set $\tilde t = t = 1$ at the outset.  We find also that the optimal $\tilde \alpha$ is always very close to $\alpha$, the difference being likely due to numerical error.  The remaining parameters are, however, significantly renormalized by interactions---$\tilde \mu$ because of a charging energy associated with $U$, and $\tilde \Delta, \tilde V_z$ for the reasons described above.  With the optimal parameters in hand, Hartree-Fock theory predicts that the system is topological when $\tilde V_z > \tilde \Delta$ and $-\sqrt{\tilde V_z^2-\tilde \Delta^2}<\tilde \mu<\sqrt{\tilde V_z^2 -\tilde \Delta^2}$, which constitutes a straightforward generalization of Eqs.\ (\ref{TopologicalCriteria}).  

\begin{figure}
\includegraphics[width=\columnwidth]{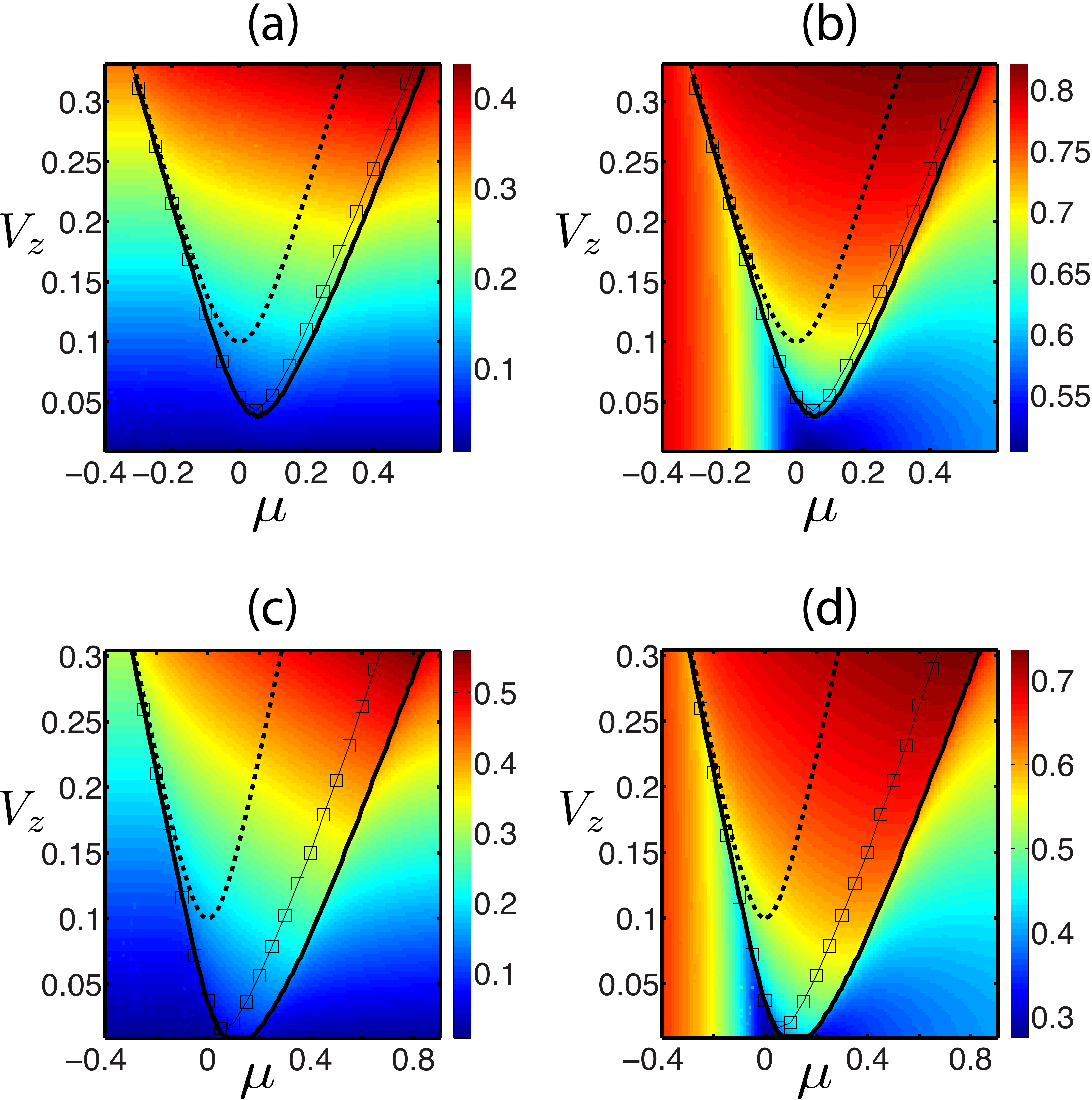}
\caption{Hartree-Fock phase diagrams with $\Delta = 0.1$, $\alpha = 0.3$, and interaction strengths $U = 0.5$ [parts (a) and (b)] and $U = 1.0$ [parts (c) and (d)].  The dashed lines represent the phase boundary separating the topological and non-topological phases when $U = 0$, while the thick solid lines represent the Hartree-Fock phase boundary for an interacting wire.  Square symbols denote DMRG data points from Fig.\ \ref{fig:phase_diagram}.  The effective Zeeman field $\tilde V_z$---which interactions enhance compared to $V_z$---is shown by the shading in (a) and (c).  The effective pairing field $\tilde \Delta$ normalized by $\Delta$ is represented by the shading in (b) and (d); this is clearly suppressed by interactions.}
\label{fig:HF}
\end{figure}

Figure \ref{fig:HF} illustrates our results when $\Delta = 0.1$ and $\alpha = 0.3$; the data in parts (a) and (b) were obtained with $U = 0.5$, while parts (c) and (d) correspond to $U = 1$.  Dashed lines indicate the topological phase boundary for the non-interacting system, square symbols denote DMRG data points from Fig.\ \ref{fig:phase_diagram}, and thick solid lines represent the Hartree-Fock phase boundary.  The shading in Figs.\ \ref{fig:HF}(a) and (c) shows the optimal $\tilde V_z$, which is enhanced throughout compared to the bare $V_z$, most strongly inside of the topological phase and for larger values of $\mu$.  Similarly, the shading in Figs.\ \ref{fig:HF}(b) and (d) indicates the optimal $\tilde \Delta$ normalized by the bare $\Delta$.  This is suppressed throughout, though most weakly when the Zeeman splitting is large.  The `dual' role of the Zeeman field underpins this trend: $V_z$ both magnetizes the system and simultaneously suppresses pairing.  Hence at large $V_z$ the anomalous correlator in $E_I$ above is already significantly suppressed, even when $\tilde \Delta = \Delta $.  The renormalization of the Zeeman and pairing fields clearly underlies the expansion of the topological phase evident in Fig.\ \ref{fig:HF}, lending further support to the interpretation provided above for our DMRG results.  

The Hartree-Fock and DMRG phase boundaries agree remarkably well in the $U = 0.5$ case, with Hartree-Fock theory only marginally overestimating the extent of the topological phase.  Poorer agreement occurs, however, when $U = 1$.  Aside from the greater quantitative overestimate of the width of the topological phase at finite $V_z$, Hartree-Fock theory fails in a more serious way as $V_z\rightarrow 0$.  Namely, as Figs.\ \ref{fig:HF}(c) and (d) indicate, the topological phase is predicted to exist over a finite range of $\mu$ even at $V_z = 0$.  This can only arise if a finite Zeeman field is \emph{spontaneously} generated at $V_z = 0$, which is forbidden by the Mermin-Wagner theorem in the model under consideration ($V_z$ breaks both time-reversal symmetry \emph{and} a continuous U(1) spin symmetry that would otherwise be present in $H$; see Sec.~\ref{sec:Vz_zero} for a more detailed discussion).

Thus while Hartree-Fock theory provides a simple, intuitive picture for the influence of interactions on the topological phase, its regime of validity is, not unexpectedly, limited.  This approach's inevitable shortcomings surface when interactions become too strong (compared to the gap in the non-interacting limit).  There the stiffness of the system against fluctuations---which is what we argued might allow Hartree-Fock theory to work in the first place---is overwhelmed by the interactions.  This limit is best attacked instead by first considering an interacting, spin-orbit coupled wire (which forms a Luttinger liquid), and \emph{then} including the effects of $\Delta$ and $V_z$.  One can pursue this route using bosonization, which was already done in Ref.\ \onlinecite{Gangadharaiah:2011} in the large-Zeeman case where the problem could be mapped onto that of spinless fermions.  In the following subsection we will consider the opposite limit where both $\Delta$ \emph{and} $V_z$ are small.

\subsection{Bosonization analysis at small $\Delta, V_z$}
\label{bosonization}
In the limit of small $\Delta$ and $V_z$ the natural starting point is a critical system of interacting electrons subject to spin-orbit 
coupling.  To describe this regime analytically we will focus now on the continuum Hamiltonian $\mathcal{H}$ in Eq.\ (\ref{eqn:continuum_H}) rather than the discrete lattice model studied in the previous two subsections.  The single-particle energy spectrum in this case is given by two horizontally shifted parabolas crossing at
$k=0$, as shown by the dashed lines in Fig.\ \ref{fig:bands}.  Recall from the discussions leading to Eqs.\ \eqref{TopologicalCriteria}
that with $\Delta = V_z = 0$ the system is closest to realizing the topological phase when $\mu = 0$.  In the non-interacting limit, perturbing this point with couplings $\Delta$ and $V_z$
such that $V_z > \Delta$ then drives the wire into a topological superconductor.  We would like to now understand how an \emph{interacting} wire with $\mu \approx 0$ enters the topological phase upon adding weak pairing and Zeeman fields $\Delta, V_z$.  

We find it convenient for the following to first unitarily rotate the Hamiltonian in Eq.\ 
\eqref{eqn:continuum_H} by $\pi/2$ about the $\sigma^x$ axis so that $\sigma^y \to -\sigma^z$ and 
$\sigma^z \to \sigma^y$.  
(This is implemented by the unitary operator $U = \exp[ - i \pi \sigma^x/4]$).  In this basis spin-up electrons with $\sigma^z=+1$ live on the right-shifted parabola and exhibit Fermi points at 
$k_{F1} = 2 m\alpha$ and $k_{F2} = 0$, while spin-down electrons belong to the left-shifted parabola
with Fermi points at momenta $-k_{F1}$ and $k_{F2}$. Low-energy excitations near these Fermi points can be captured
by decomposing the fermion operators in terms of slowly varying left- and right-moving fields $\psi_{L/R\, s}$ through
\bea
\psi_{\uparrow} &=& \psi_{R \uparrow} e^{i k_{F1} x} + \psi_{L \uparrow}
\nonumber\\
\psi_{\downarrow} &=& \psi_{R \downarrow}  + \psi_{L \downarrow} e^{-i k_{F1} x}.
\eea
The non-interacting part of the effective low-energy Hamiltonian with $V_z = \Delta = \mu = 0$ is then given by
\be
\tilde{\mathcal{H}}_0 = \sum_{s = \uparrow,\downarrow}\int_x \left[\psi^\dagger_{R s} (-i v_F \partial_x)\psi_{R s} + 
\psi^\dagger_{L s} (i v_F \partial_x)\psi_{L s}\right],
\ee
while the non-oscillatory parts of the perturbations we are interested in read
\bea
\tilde{\mathcal{H}}_{V_z} &=& \int_x i V_z \Big(\psi^\dagger_{R \downarrow} \psi_{L \uparrow} - \psi^\dagger_{L \uparrow} \psi_{R \downarrow}\Big)
\label{HVz} \\
\tilde{\mathcal{H}}_\Delta &=& \int_x \Delta \Big(\psi_{R \uparrow} \psi_{L \downarrow} + \psi_{L \uparrow} \psi_{R \downarrow} + h.c.\Big)
\label{HDelta}\\
\tilde{\mathcal{H}}_\mu &=& -\mu \sum_{s = \uparrow,\downarrow} \int_x \Big(\psi^\dagger_{R s} \psi_{R s} + \psi^\dagger_{L s} \psi_{L s}\Big) .
\label{Hmu}
\eea

We will incorporate electron-electron interactions using bosonization\cite{Giamarchi:2004}, expressing the low-energy fermion fields in terms of conjugate bosonic fields $\varphi_s, \theta_s$ as 
\be
\psi_{R s} =\frac{ e^{i\sqrt{\pi}(\varphi_s - \theta_s)}}{\sqrt{2\pi a}},~~
\psi_{L s} =\frac{ e^{-i\sqrt{\pi}(\varphi_s + \theta_s)}}{\sqrt{2\pi a}}.
\ee
We then form 
symmetric (charge) and antisymmetric (spin) combinations of the bosonic variables via
\be
\varphi_\rho = (\varphi_\uparrow + \varphi_\downarrow)/\sqrt{2}, ~~
\varphi_\sigma = (\varphi_\uparrow - \varphi_\downarrow)/\sqrt{2},
\ee
and similarly for the $\theta_s$ fields.
The {\em interacting} Hamiltonian with $V_z = \Delta = \mu = 0$ 
is represented by a sum $\tilde{\mathcal{H}} = \tilde{\mathcal{H}}_\rho + \tilde{\mathcal{H}}_\sigma$ 
for the charge and spin sectors. 
The charge sector is harmonic,
\be
\tilde{\mathcal{H}}_\rho = \frac{v_\rho}{2} \int_x \left[K_\rho (\partial_x \theta_\rho)^2 + 
\frac{1}{K_\rho} (\partial_x \varphi_\rho)^2\right],
\label{Hinteracting}
\ee
and is parameterized by a repulsive charge Luttinger parameter $K_\rho < 1$.
The spin-sector Hamiltonian contains in addition to harmonic pieces a marginally irrelevant backscattering term $\tilde{\mathcal{H}}_{\rm bs}$,
\bea
\tilde{\mathcal{H}}_\sigma &=& \frac{v_\sigma}{2} \int_x \left[K_\sigma (\partial_x \theta_\sigma)^2 + \frac{1}{K_\sigma} (\partial_x \varphi_\sigma)^2\right] + \tilde{\mathcal{H}}_{\rm bs},\nonumber\\
\tilde{\mathcal{H}}_{\rm bs} &=& - g_{\rm bs}\int_x {\bf J}_R \cdot {\bf J}_L .
\label{Hint-sigma}
\eea
While a bosonized expression of $\tilde{\mathcal{H}}_{\rm bs}$ is available, we find it convenient 
to formulate this interaction in terms of spin currents 
$J^a_{R/L} = \frac{1}{2}\psi^\dagger_{R/L}\sigma^a \psi_{R/L}$ describing spin-density fluctuations 
at the right/left Fermi points. In this representation the harmonic part of $\tilde{\mathcal{H}}_\sigma$
describes the spin sector for non-interacting electrons, with all interaction terms for this sector encoded in $\tilde{\mathcal{H}}_{\rm bs}$.  
For weak interactions the backscattering amplitude is given by $g_{\rm bs} \sim 2U$.
Also, for both sectors the product $v_\nu K_\nu = v_F$ remains unrenormalized.

A brief digression is in order regarding the spin backscattering term in Eq.\ (\ref{Hint-sigma}) and the charge Luttinger parameter $K_\rho$.  Naively, one might expect that the SU(2)-invariant form of $\tilde{\mathcal{H}}_{\rm bs}$ above is not appropriate for our system because of the presence of spin-orbit coupling.  However, this is not the case for the specific models we have been considering, which in fact exhibit a `hidden' SU(2) symmetry that guarantees the legitimacy of Eq.\ (\ref{Hint-sigma}).  This symmetry is most simply exposed in the continuum Hamiltonian in Eq.\ (\ref{eqn:continuum_H}) when $V_z = \Delta = 0$ by defining new operators 
$\psi = \text{exp}[-i m \alpha\,\sigma^y\, x]\, \psi'$.  When the Hamiltonian is expressed in terms of $\psi'$, the spin-orbit term is absent while the interaction term retains exactly the same form---\emph{i.e.}, the transformed Hamiltonian exhibits SU(2) symmetry.  A similar conclusion applies for the lattice model in Eq.\ (\ref{H}), which can be mapped onto the pure Hubbard model with renormalized hopping $t' = \sqrt{t^2 + \alpha^2}$ upon performing a suitable unitary transformation to eliminate the spin-orbit term.\cite{Shekhtman:1992}  It also follows that, just as in the pure Hubbard model, the charge Luttinger parameter appropriate for our DMRG simulations satisfies\cite{LuttingerParameterHubbardModel} $1/2 \leq K_\rho \leq 1$, even when $\alpha \neq 0$.  

In bosonized language the perturbations in Eqs.\ (\ref{HVz}) through (\ref{Hmu}) read
\bea
\label{pert:V}
\tilde{\mathcal{H}}_{V_z} &=& \int_x  \frac{V_z}{\pi a} \sin[\sqrt{2\pi}(\varphi_\rho + \theta_\sigma)],\\
\label{pert:delta}
\tilde{\mathcal{H}}_\Delta &=& \int_x \frac{2\Delta}{\pi a} \cos[\sqrt{2\pi} \theta_\rho] \cos[\sqrt{2\pi} \varphi_\sigma], \\
\label{pert:mu}
\tilde{\mathcal{H}}_\mu &=& -\mu \int_x  \sqrt{\frac{2}{\pi}} ~\partial_x \varphi_\rho .
\eea
Observe that $\tilde{\mathcal{H}}_{V_z}$ and $\tilde{\mathcal{H}}_\Delta$ are expressed in terms of {\em dual} fields relative to one another:
$\varphi_\rho$ vs. $\theta_\rho$ and $\theta_\sigma$ vs. $\varphi_\sigma$ correspondingly.
The scaling dimensions of the perturbing operators are
\bea
D_{V_z} &=& \frac{1}{2}\left(K_\rho + \frac{1}{K_\sigma}\right) = \frac{1}{2}(1 + K_\rho)\\
D_\Delta &=& \frac{1}{2}\left(\frac{1}{K_\rho} + K_\sigma\right) =  \frac{1}{2}\left(1 + \frac{1}{K_\rho}\right)\\
D_\mu &=& 1.
\eea
Under renormalization the corresponding couplings grow according to
\bea
\frac{d V_z}{d\ell} &=& \frac{(3-K_\rho)}{2} V_z + \frac{1}{4} \frac{g_{\rm bs}}{2\pi v_F} V_z,\nonumber\\
\frac{d \Delta}{d\ell} &=& \frac{(3-K_\rho^{-1})}{2} \Delta - \frac{3}{4} \frac{g_{\rm bs}}{2\pi v_F} \Delta,\nonumber\\
\frac{d \mu}{d\ell} &=& \mu,
\label{RGeqns}
\eea
where $\ell$ is a logarithmic rescaling factor.  The contributions to the flow equations containing the backscattering coupling $g_{\rm bs}$ are most conveniently derived using the original fermionic representations in Eqs.\ \eqref{HVz}, \eqref{HDelta}
and \eqref{Hint-sigma}.  One can perturbatively expand the action in these terms and then apply Wick's theorem 
to fuse fermions at nearby space-time points.  Key technical details of this procedure can be found, for example,
in Ref.\ \onlinecite{starykh05}.  Alternatively, a standard one-loop momentum-shell renormalization group analysis leads to the same results.  It is worth noting that in our calculation the renormalization group flow of $g_{\rm bs}$ is standard,
\begin{equation}
  \frac{d g_{\rm bs}}{d\ell} = - \frac{g_{\rm bs}^2}{2\pi v_F},
\end{equation}
so that $g_{\rm bs}(\ell)/(2\pi v_F) = g_0/(2\pi v_F + g_0 \ell)$.
Here $g_0$ denotes initial value of the backscattering interaction. Note also that $\mu$, which is expressed
in terms of charge fields only in Eq.\ \eqref{pert:mu}, is completely unaffected by $\tilde{\mathcal{H}}_{\rm bs}$.

The renormalization group equations \eqref{RGeqns} are solved by 
\bea
V_z(\ell) &=& V_z \exp\left[\frac{(3-K_\rho)}{2}\ell\right] \Big(1 + \frac{g_0 \ell}{2\pi v_F}\Big)^{1/4}, \nonumber \\
\Delta(\ell) &=& \Delta \exp\left[\frac{(3-K_\rho^{-1})}{2}\ell\right] \Big(1 + \frac{g_0 \ell}{2\pi v_F}\Big)^{-3/4},\nonumber \\
\mu(\ell) &=& \mu e^\ell,
\label{RGsol}
\eea
where $V_z, \Delta, \mu$ on the right-hand-side stand for initial values of these couplings. 
Thus $V_z$ and $\mu$ constitute relevant perturbations for any repulsive interaction strength, while $\Delta$ becomes irrelevant when $K_\rho < 1/3$.  In what follows we will assume $1/3 < K_\rho < 1$ so that all three couplings are relevant.  (For $K_\rho < 1/3$ interactions are too strong for superconductivity---topological or otherwise---to take root.)  

Let us first explore the most important case of $\mu=0$, where the competition between the
Zeeman and superconducting perturbations is most apparent.  Since both are relevant in the parameter range of interest, the fate of the system depends on which of the two renormalized parameters first flows to strong coupling (\emph{i.e.}, values of order the Fermi velocity $v_F$).  If the initial conditions are such that $V_z$ dominates, then the topological superconducting state emerges, whereas if $\Delta$ dominates an ordinary superconducting state appears.  Comparing the renormalization group scales $\ell_{V_z}$ and $\ell_\Delta$
at which the renormalized parameters reach strong coupling, we find that the topological phase arises provided
\be
V_z > c\,\Delta^\beta ~\text{with} 
~\beta \approx \frac{3 - K_\rho + g_0/(4\pi v_F)}{3 - K_\rho^{-1} - 3g_0/(4\pi v_F)} 
\label{eqn:curvature}
\ee
for some constant $c$.  Observe that for repulsive interactions the exponent $\beta > 1$,
resulting in a convex phase boundary in the $\Delta - V_z$ plane.  Consequently, interactions allow one to access the topological phase at weaker Zeeman fields relative to the non-interacting case, which is qualitatively consistent with what we found earlier in our DMRG and Hartree-Fock studies at finite $\Delta$.  Note that in deriving Eq.\ \eqref{eqn:curvature} we
assumed $g_0 \ll v_F$; in this regime backscattering-induced corrections to $V_z$ and $\Delta$
change little in comparison with the exponential terms in Eqs.\ \eqref{RGsol}.  It is interesting that despite the marginal irrelevance of the backscattering, at least 
in the weak coupling regime, where $1 - K_\rho \sim U/(\pi v_F)$, the spin sector contribution to the exponent $\beta$
is in fact comparable to that of the charge sector. 

Similar considerations for the competition between $V_z$ and $\mu$ imply that at weak 
$\Delta$ the topological phase appears when
\be
V_z > c'|\mu|^{\beta'}~ \text{with} ~ \beta' \approx (3 - K_\rho + g_0/(4\pi v_F))/2,
\ee
where $c'$ is a constant.  
Since $\beta' > 1$, the chemical potential window over which this state occurs broadens, 
which is again in line with our earlier numerical results.  

As an aside, we briefly contrast our results with the predictions of Gangadharaiah \emph{et al}.\cite{Gangadharaiah:2011} for the large-Zeeman limit, where only two rather than four fields are needed to capture the low-energy physics.  We can access their limit from our approach by first observing that at large $V_z$ the combination $\phi_\rho + \theta_\sigma$ is pinned by $\tilde{\mathcal{H}}_{V_z}$, reducing the number of gapless modes by a factor of two.  The remaining gapless mode is described by fields $\theta = (\theta_\rho-\phi_\sigma)/\sqrt{2}$ and $\phi = (\phi_\rho-\theta_\sigma)/\sqrt{2}$.  Integrating out massive fluctuations and rewriting $\tilde{\mathcal{H}}$ in terms of $\theta,\phi$, one obtains
\begin{equation}
  \tilde{\mathcal{H}}_{{\rm large} ~V_z} =  \frac{1}{2}\int_x ~v \left[K (\partial_x \theta)^2 + \frac{1}{K} (\partial_x \varphi)^2\right],
\end{equation}
with 
\begin{eqnarray}
  K &=& \frac{2}{\sqrt{K_\sigma^2 + (v_\sigma K_\sigma)/(v_\rho K_\rho) + (v_\rho/K_\rho)(K_\sigma/v_\sigma) + K_\rho^{-2}}}
  \nonumber \\
  &=& \frac{2}{\sqrt{(1+K_\sigma^2)(1+K_\rho^{-2})}}.
  \label{KlargeVz}
\end{eqnarray}
(For simplicity, we neglected the marginally-irrelevant backscattering term here.)  At the fixed point with only one gapless mode, the pairing term becomes irrelevant for $K< 1/2$.\cite{Gangadharaiah:2011}  It is tempting to try use Eq.\ (\ref{KlargeVz}) to reconcile this result with our finding that when $V_z \rightarrow 0$ and $K_\sigma = 1$ pairing becomes irrelevant when $K_\rho < 1/3$.  Such a comparison is difficult, however, since the Luttinger parameters $K_\rho$ and $K_\sigma$ will flow away from their $V_z = 0$ values as the Zeeman field renormalizes to strong coupling.  We simply remark that in general there is no reason to expect the critical interaction strength above which pairing becomes irrelevant to be the same at weak and strong $V_z$.  This opens the intriguing possibility that there exists a range of interactions where the topological phase can be accessed at weak $V_z$, but beyond a critical value is replaced by a gapless state.  Such a scenario is very different from the case of non-interacting electrons where the topological phase can in principle appear at arbitrarily large $V_z$, and would be interesting to explore in future numerical work.  

\subsection{DMRG study of the phase diagram with interactions II}
\label{DMRGII}

\begin{figure}[tp]
\includegraphics[width=\columnwidth]{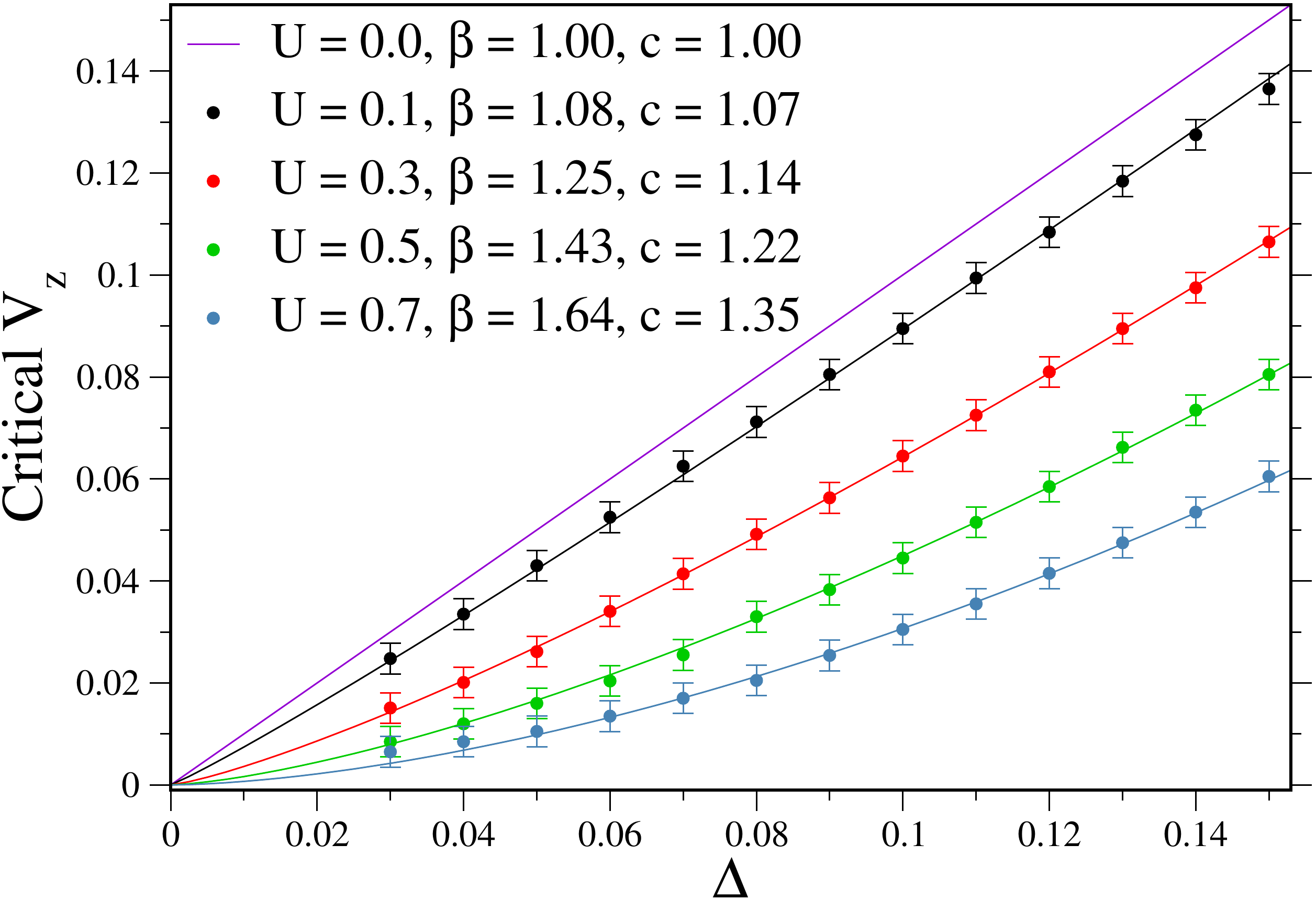}
\caption{Critical $V_z$ values separating the topological phase ($V_z>V_z^{\text{crit}}$) from the conventional phase ($V_z<V_z^{\text{crit}}$) . The DMRG data
was obtained by scanning $V_z$ at fixed $\Delta$ for an 800 site system with $\alpha=0.3$. We have chosen $\mu$ according to the relation $\mu = (0.0832)U$, which we find empirically cancels the charging effect arising from the on-site repulsion and keeps the density fixed at its $U=0$, $\mu=0$ value.  The solid lines are fits of the form $c\,\Delta^\beta$ with the exponents $\beta$ evaluated from Eq.~(\ref{eqn:curvature}) taking $g_0 = U/2$ and $K_\rho$ from the Bethe ansatz. }
\label{fig:phase_boundary}
\end{figure}

To verify our perturbative bosonization predictions and test their range of applicability
we will now quantitatively compare these results with DMRG data.  Doing so requires us to extend our DMRG simulations to smaller values of $\Delta$ than we analyzed previously.  To minimize finite-size effects caused by long correlation lengths in this regime, our numerics for this section were performed on an 800-site system which is twice as large as those previously simulated.  
The data points in Fig.\ \ref{fig:phase_boundary} illustrate the computed phase boundary separating the topological and conventional
phases in the $\Delta$ -- $V_z$ plane, with $\alpha = 0.3$ and at several values of $U$. 
Qualitative agreement with Eq.~(\ref{eqn:curvature}) is immediately apparent;
interactions indeed cause the topological phase to broaden and occur at weaker values of $V_z$.

For a more detailed comparison, we fit the critical $V_z$ values computed with DMRG to a power law of the form 
$V_z^{\text{crit}}(\Delta) = c\,\Delta^\beta$.  Treating both $c$ and $\beta$ as fitting parameters indeed allows one to fit \emph{all} of the data from Fig.\ \ref{fig:phase_boundary} extremely well (not shown), even those at the largest $\Delta$ considered where the validity of our bosonization analysis is not \emph{a priori} obvious.  It is interesting to ask whether $\beta$ can be calculated rather than treated as a fitting parameter, since $K_\rho$ can be obtained exactly for the pure Hubbard model.\cite{LuttingerParameterHubbardModel}  Such an analysis is complicated, however, by the influence of the spin-sector backscattering term on the exponent $\beta$, particularly given that all of the $U>0$ curves are rather far from the regime where this term may be treated perturbatively. 

It is instructive to first attempt to fit the data by computing $\beta$ with $g_0$ artificially set to zero in Eq.\ (\ref{eqn:curvature}).  To obtain $K_\rho$, we first perform the unitary transformation discussed in the previous section to map the $V_z = \Delta = 0$ lattice Hamiltonian onto the pure Hubbard model with hopping parameter $t' = \sqrt{t^2 + \alpha^2}$.  
One can then numerically solve the Bethe ansatz equations appearing, for example, in 
Ref.~\onlinecite{Lieb:1968}, and determine the charge-sector velocity and Luttinger parameter by calculating both the compressibility and the response of the system
to a small flux.\cite{LuttingerParameterHubbardModel,Giamarchi:2004}
[Note that an interaction of strength $U$ in our DMRG simulations corresponds to $2U$ in the usual Hubbard model literature because of the factor of 1/2 multiplying the hopping term in Eq.\ (\ref{H0}).]  When the exponent $\beta$ is determined in this fashion, we find that the fit to our DMRG data is in fact rather poor.  This strongly suggests that the spin backscattering term---despite being marginally irrelevant---in fact plays an important quantitative role, as argued in the previous section.  

It is unfortunately difficult to obtain a reliable quantitative estimate for the initial backscattering coupling $g_0$ outside of the perturbative regime (which does not apply here).  Thus we will instead assume an ansatz $g_0 = b U$, where $b$ is a fitting parameter independent of $U$.  Note that this is far more restrictive than treating $\beta$ as a separate fitting parameter for each value of $U$.  By computing $\beta$ from Eq.~(\ref{eqn:curvature}) and treating $b$ and $c$ as a fitting parameters,
we obtain the fits shown by solid lines in Fig.~\ref{fig:phase_boundary}.  These curves correspond to $b = 1/2$ and values of $c$ and $\beta$ listed in the figure.
The agreement between field theoretic predictions 
and DMRG is rather remarkable, and holds for all $\Delta$ and $U$ shown.  
This agreement is particularly encouraging since the two methods compared here are complementary in the sense that DMRG operates optimally with short correlation
lengths while bosonization works perturbatively in small $\Delta$ and $V_z$ where correlation lengths are long. 
By applying both methods one can seamlessly connect these two very different limits.

\subsection{Topological superconductivity at zero magnetic field \label{sec:Vz_zero}}

The results from the preceding subsections collectively paint the following general picture: local Coulomb repulsion can significantly broaden the topological superconducting state in phase space, allowing the topological regime to be accessed at smaller Zeeman fields and over a wider chemical potential window.  Given the trend exposed by our DMRG data in Fig.\ \ref{fig:phase_diagram}, it is natural to ask whether, at still larger interaction strength $U$, the topological phase might extend all the way down to $V_z = 0$.  We saw in Sec.\ \ref{sec:HF} that Hartree-Fock theory in fact makes such a prediction, but unfortunately dismissed this as an unphysical artifact on very general grounds.  

Let us discuss the underlying reason in greater detail here.  Recall that realizing the topological phase requires the wire to enter a `spinless' regime where there exists only a single set of Fermi points.  Due to Kramer's theorem, this is fundamentally impossible in a time-reversal-invariant system.  One can understand this in the context of the band structure in Fig.\ \ref{fig:bands} by observing that the $k = 0$ crossing of the dashed lines (which correspond to $V_z = \Delta = 0$) is protected by time-reversal symmetry.  The applied Zeeman field serves the sole purpose of breaking this symmetry, but as an inessential byproduct simultaneously breaks a second, \emph{continuous} symmetry exhibited by the Hamiltonians we have been considering up to this point.  Namely, when $V_z = 0$ the Hamiltonian $H$ in Eq.\ (\ref{H}) (like its continuum version $\mathcal{H}$) is invariant under global U(1) spin rotations about the $y$-axis, which is lifted when $V_z \neq 0$.  In principle, time-reversal symmetry---which is discrete---\emph{can} be broken spontaneously, though this U(1) symmetry cannot.  

We now pose the following question: how generic is the continuous U(1) symmetry which prevents the spontaneous generation of a Zeeman field?  For the strongly spin-orbit coupled wires we have implicitly been considering throughout, this U(1) is in fact quite far from being a microscopic symmetry.  Its presence merely reflects our inclusion of substrate-induced Rashba coupling only, and our neglect of Dresselhaus spin-orbit terms that are intrinsic to the wire.  A more realistic model which accounts for both types of spin-orbit interactions will generically exhibit, at most, time-reversal and the discrete space-group symmetries exhibited by the particular wire under consideration.  The barrier to generating a Zeeman field spontaneously through strong interactions is then lifted, in principle allowing the topological phase to be accessed \emph{without} the explicit application of a magnetic field.  

\begin{figure}[tp]
\includegraphics[width=\columnwidth]{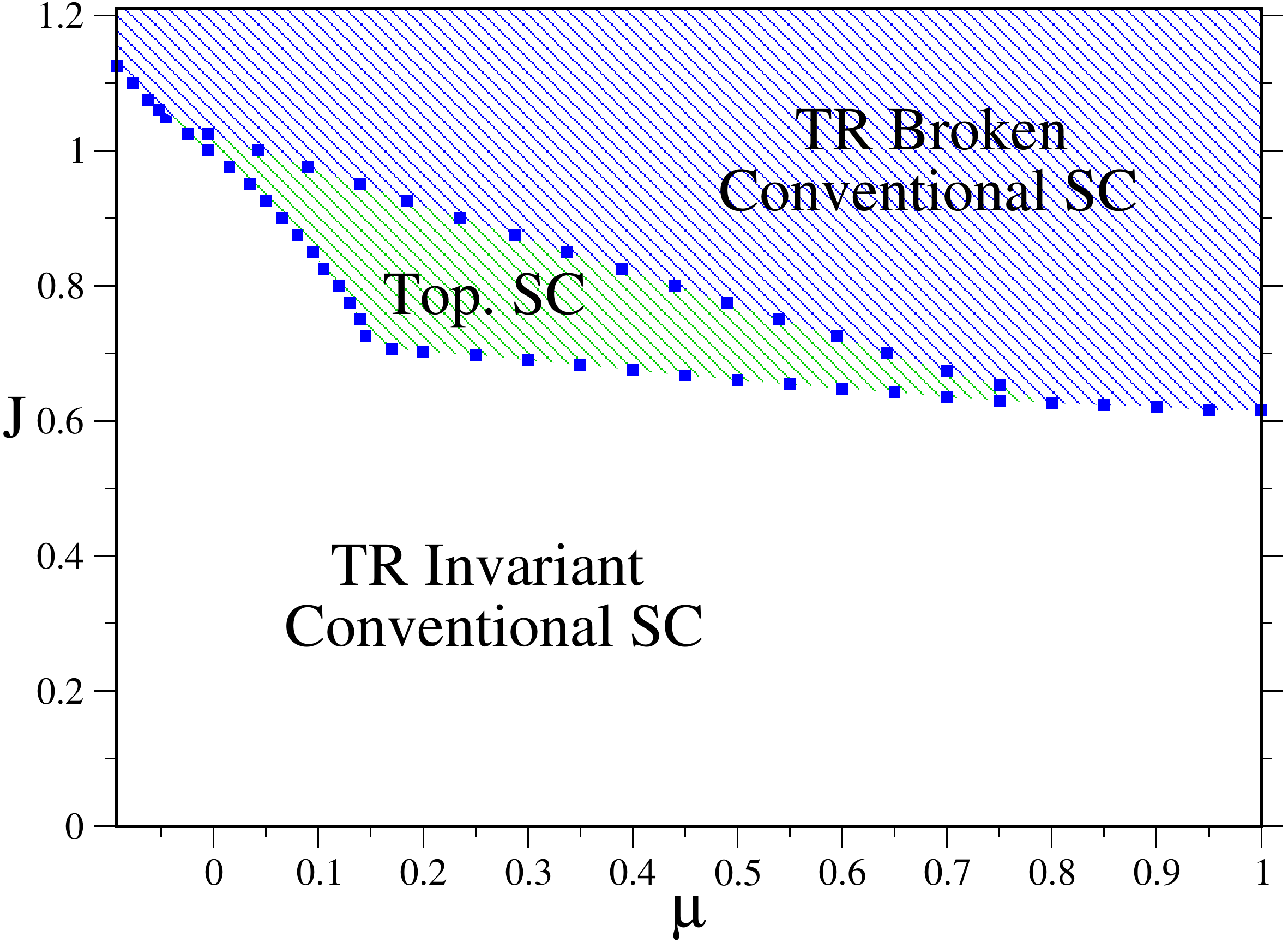}
\caption{Phase diagram of a wire with spin-orbit coupling $\alpha=0.3$, proximity-induced pairing $\Delta=0.1$,
and Ising-like interactions given by Eq.~(\ref{HIprime}). Because of the first-order nature of the phase transitions, reliable DMRG
data could be obtained for systems as small as 100 sites, though larger systems were also used to check results. 
Error bars are represented by the symbol sizes.
By using the same `order parameters' discussed in Sec.~\ref{sec:DMRG_Study_I}, we observe a spontaneously generated
topological superconducting phase over a large region of the phase diagram \emph{without} an applied Zeeman field.}
\label{fig:spontaneous_pd}
\end{figure}

As proof of this principle, we use DMRG to study a lattice model where the U(1) symmetry is broken not by additional spin-orbit terms, but rather through the interactions themselves:
\begin{eqnarray}
  H' &=& H_0' + H_I'
  \label{Hprime}
  \\
  H_0' &=& \sum_j \bigg{[}-\frac{t}{2} (c_j^\dagger c_{j+1} + h.c.) -(\mu-t)c_j^\dagger c_j 
  \label{H0prime}
  \nonumber \\
  &-& \frac{\alpha}{2}(i c_j^\dagger \sigma^y c_{j+1} + h.c.)
  + \Delta(c_{j\uparrow}c_{j\downarrow} + h.c.)\bigg{]}
  \\
  H_I' &=& -J\sum_j (c_j^\dagger \sigma^z c_j)(c_{j+1}^\dagger \sigma^z c_{j+1}), \label{HIprime}
\end{eqnarray}
with $J> 0$.  Here $H_0'$ describes a Rashba-coupled wire with proximity-induced pairing, but without a Zeeman field.  
The interaction term $H_I'$ favors ferromagnetically aligning the spins either along or against the $z$ direction, thereby lifting the U(1) spin symmetry exhibited by $H_0'$.  Note that the entire Hamiltonian is time-reversal symmetric and possesses no continuous symmetries.

\begin{figure}
\includegraphics[width=\columnwidth]{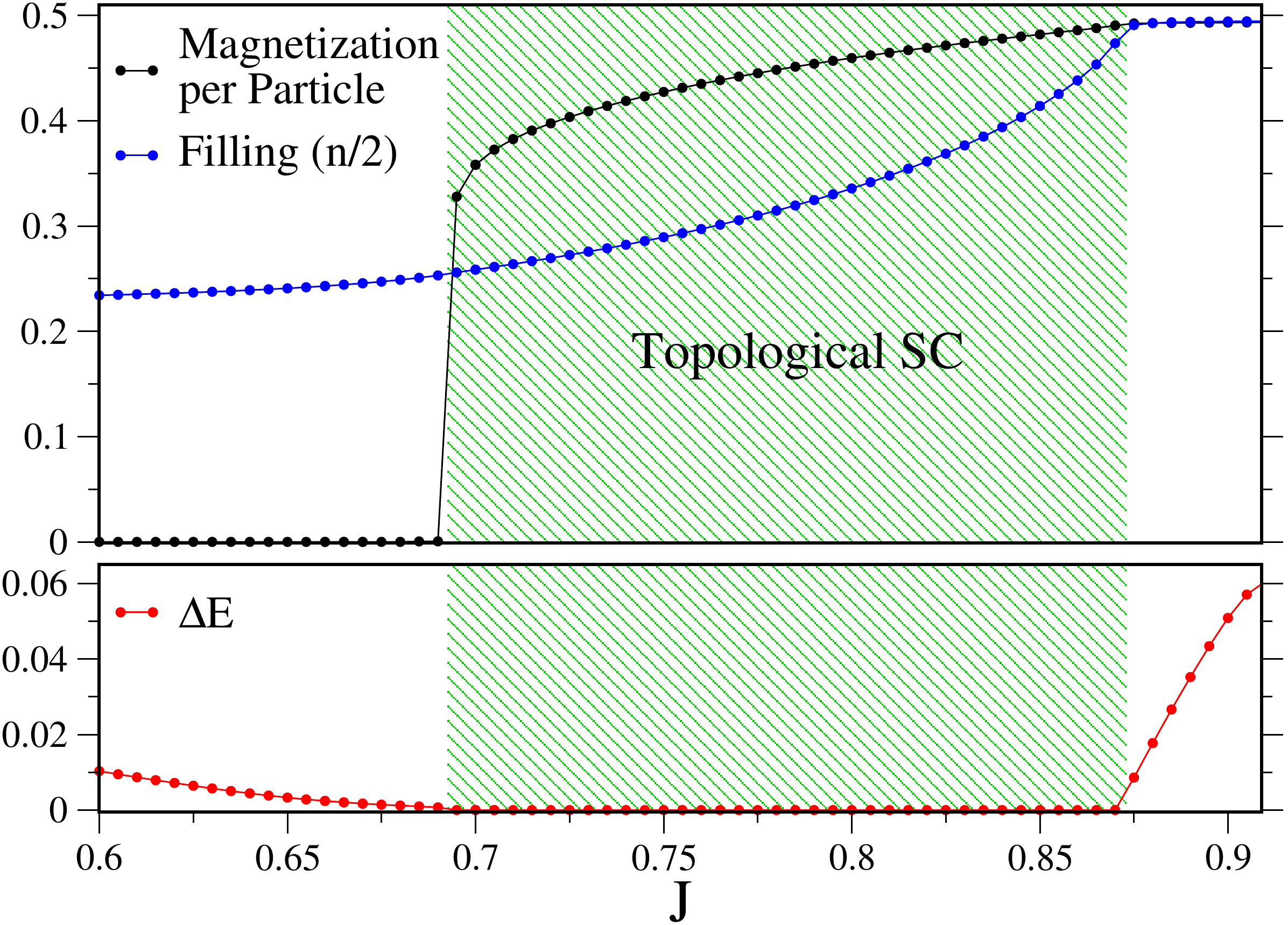}
\caption{Cut through the phase diagram of Fig.~\ref{fig:spontaneous_pd} at fixed $\mu=0.3$. 
Once $J$ increases to $J\simeq 0.693$ the magnetization undergoes a strong first-order jump to a finite value and the system enters
the topological phase signaled by $\Delta E = 0$. At the upper boundary $J\simeq 0.872$ the magnetization and site occupancy both saturate to a value
of nearly 1/2 and the system undergoes a transition to a conventional time-reversal breaking superconducting phase.}
\label{fig:cut}
\end{figure}

Before turning to DMRG to determine this model's phase diagram, it is useful to consider some simple limits with $\alpha, \Delta$ finite.
In the non-interacting case $J = 0$, the absence of a Zeeman field implies that the Hamiltonian can only capture an ordinary, non-topological superconducting phase.  Since this phase is protected by a finite energy gap it must persist up to some finite value of $J$.  In the opposite limit where $J\rightarrow\infty$, the model reduces to a ferromagnetic Ising chain which spontaneously breaks time-reversal symmetry by singly occupying every site with perfectly spin-polarized electrons.  For $J$ large (compared to all energy scales) but finite, spin-orbit coupling leads to imperfect spin-polarization and the pairing field then renders the system weakly superconducting.  Though time-reversal symmetry breaking still survives here, this superconducting state is nevertheless non-topological because there are no partially filled bands.  

The spontaneous onset of topological superconductivity requires both time-reversal symmetry breaking \emph{and} a particular window of non-interacting
electron densities.  
While it is far from obvious that these conditions are ever met in this model, our DMRG results presented
in Fig.~\ref{fig:spontaneous_pd} reveal that topological superconductivity exists over a significant region of the phase diagram.
In Fig.~\ref{fig:cut} we show the characteristic behavior of the system along a cut through the phase diagram at fixed $\mu=0.3$.
As $J$ increases from zero, the system undergoes a strongly first-order transition beyond which the electron spins align ferromagnetically;
 $\Delta E$ simultaneously drops to zero signaling the onset of the topological phase. At the upper boundary, the magnetization saturates to a value 
that is close to full spin polarization and the system becomes nearly half filled.

Though we do not expect the Ising interaction Eq.~(\ref{HIprime}) to arise in an actual experimental wire, 
we take the existence of topological order in this model as motivation for future investigation of 
more realistic models that may spontanously enter a topological phase. For instance, one might consider additional 
spin-orbit terms that break the problematic U(1) spin symmetry together with strong Coulomb interactions. We leave
such considerations for a future study.

\section{Discussion}

Interacting wires proximate to a bulk $s$-wave superconductor provide an experimentally relevant research area in which Coulomb repulsion, magnetism, and Cooper pairing coexist and compete in interesting ways.  Moreover, because the low-energy degrees of freedom reside in one dimension, the problem is amenable to powerful numerical and analytical approaches including DMRG and bosonization, and interestingly can even be attacked reasonably well using far cruder Hartree-Fock theory when the proximity effect and applied magnetic fields are sufficiently strong. 

Here we employed all three of these methods, focusing on how on-site repulsion affects the onset of topological superconductivity in the wire.  A rather intuitive picture emerged from these studies---interactions suppress pairing and hence the bulk gap protecting the topological phase.  As a corollary, the localization length of the Majorana modes in this state increases with interactions.  While these features are undesirable from an experimental standpoint, there is an important silver lining.  Namely, interactions additionally enhance the magnetization in the wire, thereby broadening significantly the chemical potential window over which the topological phase emerges.  This implies a greater robustness of the topological phase against chemical potential fluctuations.  In our view this boon more than compensates for the decrease in the bulk gap, since it is likely that---at least initially---disorder in the wire will present the main obstacle to be overcome as opposed to low-temperature requirements.  We further proposed that in wires with appreciable Rashba \emph{and} Dresselhaus coupling, strong repulsive interactions may eliminate the need for applied magnetic fields altogether.  As proof of principle, we demonstrated using DMRG the emergence of topological superconductivity in a strongly interacting, \emph{time-reversal invariant} toy model.  

Many interesting future directions remain to be explored.  For instance, including longer-range repulsion in DMRG simulations will allow one to access the regime in which the charge Luttinger parameter $K_\rho$ drops below 1/2.  As discussed in Ref.\ \onlinecite{Gangadharaiah:2011} and Sec.\ \ref{bosonization}, here interactions can have qualitatively new effects---notably, the complete destruction of gapped superconducting states---that are inaccessible with only on-site repulsion.  Studying the crossover between the weak- and strong-Zeeman limits where the condition for destroying superconductivity can be rather different would be particularly interesting.  Extending such simulations to incorporate both Rashba and Dresselhaus coupling is another worthwhile direction to pursue, and may allow one to find more realistic models where strong interactions lead to the formation of Majorana fermions without the presence of magnetic fields.  Multi-channel wires, which have received considerable attention recently,\cite{WimmerMultichannel,Multichannel1,Multichannel2,Multichannel3,Multichannel4,Lutchyn:2011} can also be efficiently studied with DMRG by considering multi-leg ladders.  Finally, a conceptually very interesting problem relevant for cold-atom realizations of one-dimensional topological superconductivity\cite{ColdAtomMajoranas} is the prospect of seeing remnants of Majorana physics in an \emph{attractively} interacting wire, without a proximate superconductor.  This poses quite a subtle problem because here pairing correlations will die off as a power law, but should nevertheless be tractable using DMRG.

\acknowledgments{It is a pleasure to acknowledge helpful discussions with Lukasz Fidkowski, Roman Lutchyn, Roderich Moessner, Lesik Motrunich
and Felix von Oppen. 
We would especially like to thank Steve White for not only contributing helpful discussions but also DMRG code and methods 
without which this paper would not be possible.
We gratefully acknowledge support from the National Science Foundation through grants DMR-0907500 (E.M.S.), DMR-1055522 (J.A.), DMR-0808842 (O.A.S.), and DMR-0529399 (M.P.A.F.). This research was supported in part by the National Science Foundation under Grant No. NSF PHY05-51164. O.A.S. also acknowledges the KITP, where part of this work was carried out, for hospitality. }

\bibliography{topological_wires}

\begin{thebibliography}{54}
\expandafter\ifx\csname natexlab\endcsname\relax\def\natexlab#1{#1}\fi
\expandafter\ifx\csname bibnamefont\endcsname\relax
  \def\bibnamefont#1{#1}\fi
\expandafter\ifx\csname bibfnamefont\endcsname\relax
  \def\bibfnamefont#1{#1}\fi
\expandafter\ifx\csname citenamefont\endcsname\relax
  \def\citenamefont#1{#1}\fi
\expandafter\ifx\csname url\endcsname\relax
  \def\url#1{\texttt{#1}}\fi
\expandafter\ifx\csname urlprefix\endcsname\relax\def\urlprefix{URL }\fi
\providecommand{\bibinfo}[2]{#2}
\providecommand{\eprint}[2][]{\url{#2}}

\bibitem[{\citenamefont{Wilczek}(2009)}]{WilczekPerspective}
\bibinfo{author}{\bibfnamefont{F.}~\bibnamefont{Wilczek}},
  \bibinfo{journal}{Nature Physics} \textbf{\bibinfo{volume}{5}},
  \bibinfo{pages}{614} (\bibinfo{year}{2009}).

\bibitem[{\citenamefont{Franz}(2010)}]{FranzPerspective}
\bibinfo{author}{\bibfnamefont{M.}~\bibnamefont{Franz}},
  \bibinfo{journal}{Physics} \textbf{\bibinfo{volume}{3}}, \bibinfo{pages}{24}
  (\bibinfo{year}{2010}).

\bibitem[{\citenamefont{Stern}(2010)}]{SternPerspective}
\bibinfo{author}{\bibfnamefont{A.}~\bibnamefont{Stern}},
  \bibinfo{journal}{Nature} \textbf{\bibinfo{volume}{464}},
  \bibinfo{pages}{187} (\bibinfo{year}{2010}).

\bibitem[{\citenamefont{Service}(2011)}]{Service:2011}
\bibinfo{author}{\bibfnamefont{R.~F.} \bibnamefont{Service}},
  \bibinfo{journal}{Science} \textbf{\bibinfo{volume}{332}},
  \bibinfo{pages}{193} (\bibinfo{year}{2011}).

\bibitem[{\citenamefont{Read and Green}(2000)}]{ReadGreen}
\bibinfo{author}{\bibfnamefont{N.}~\bibnamefont{Read}} \bibnamefont{and}
  \bibinfo{author}{\bibfnamefont{D.}~\bibnamefont{Green}},
  \bibinfo{journal}{Phys.\ Rev.\ B} \textbf{\bibinfo{volume}{61}},
  \bibinfo{pages}{10267} (\bibinfo{year}{2000}).

\bibitem[{\citenamefont{{Das Sarma} et~al.}(2006)\citenamefont{{Das Sarma},
  Nayak, and Tewari}}]{SrRu}
\bibinfo{author}{\bibfnamefont{S.}~\bibnamefont{{Das Sarma}}},
  \bibinfo{author}{\bibfnamefont{C.}~\bibnamefont{Nayak}}, \bibnamefont{and}
  \bibinfo{author}{\bibfnamefont{S.}~\bibnamefont{Tewari}},
  \bibinfo{journal}{Phys.\ Rev.\ B} \textbf{\bibinfo{volume}{73}},
  \bibinfo{pages}{220502(R)} (\bibinfo{year}{2006}).

\bibitem[{\citenamefont{Fu and Kane}(2008)}]{FuKane}
\bibinfo{author}{\bibfnamefont{L.}~\bibnamefont{Fu}} \bibnamefont{and}
  \bibinfo{author}{\bibfnamefont{C.~L.} \bibnamefont{Kane}},
  \bibinfo{journal}{Phys.\ Rev.\ Lett.} \textbf{\bibinfo{volume}{100}},
  \bibinfo{pages}{096407} (\bibinfo{year}{2008}).

\bibitem[{\citenamefont{Fu and Kane}(2009)}]{MajoranaQSHedge}
\bibinfo{author}{\bibfnamefont{L.}~\bibnamefont{Fu}} \bibnamefont{and}
  \bibinfo{author}{\bibfnamefont{C.~L.} \bibnamefont{Kane}},
  \bibinfo{journal}{Phys.\ Rev.\ B} \textbf{\bibinfo{volume}{79}},
  \bibinfo{pages}{161408(R)} (\bibinfo{year}{2009}).

\bibitem[{\citenamefont{Sato et~al.}(2009)\citenamefont{Sato, Takahashi, and
  Fujimoto}}]{FujimotoColdAtoms}
\bibinfo{author}{\bibfnamefont{M.}~\bibnamefont{Sato}},
  \bibinfo{author}{\bibfnamefont{Y.}~\bibnamefont{Takahashi}},
  \bibnamefont{and} \bibinfo{author}{\bibfnamefont{S.}~\bibnamefont{Fujimoto}},
  \bibinfo{journal}{Phys.\ Rev.\ Lett.} \textbf{\bibinfo{volume}{103}},
  \bibinfo{pages}{020401} (\bibinfo{year}{2009}).

\bibitem[{\citenamefont{Sato and Fujimoto}(2009)}]{SatoFujimoto}
\bibinfo{author}{\bibfnamefont{M.}~\bibnamefont{Sato}} \bibnamefont{and}
  \bibinfo{author}{\bibfnamefont{S.}~\bibnamefont{Fujimoto}},
  \bibinfo{journal}{Phys.\ Rev.\ B} \textbf{\bibinfo{volume}{79}},
  \bibinfo{pages}{094504} (\bibinfo{year}{2009}).

\bibitem[{\citenamefont{Sau et~al.}(2010{\natexlab{a}})\citenamefont{Sau,
  Lutchyn, Tewari, and {Das Sarma}}}]{Sau}
\bibinfo{author}{\bibfnamefont{J.~D.} \bibnamefont{Sau}},
  \bibinfo{author}{\bibfnamefont{R.~M.} \bibnamefont{Lutchyn}},
  \bibinfo{author}{\bibfnamefont{S.}~\bibnamefont{Tewari}}, \bibnamefont{and}
  \bibinfo{author}{\bibfnamefont{S.}~\bibnamefont{{Das Sarma}}},
  \bibinfo{journal}{Phys.\ Rev.\ Lett.} \textbf{\bibinfo{volume}{104}},
  \bibinfo{pages}{040502} (\bibinfo{year}{2010}{\natexlab{a}}).

\bibitem[{\citenamefont{Lee}(unpublished)}]{PatrickProposal}
\bibinfo{author}{\bibfnamefont{P.~A.} \bibnamefont{Lee}},
  \bibinfo{journal}{arXiv:0907.2681}  (\bibinfo{year}{unpublished}).

\bibitem[{\citenamefont{Linder et~al.}(2010)\citenamefont{Linder, Tanaka,
  Yokoyama, Sudb\o{}, and Nagaosa}}]{Linder}
\bibinfo{author}{\bibfnamefont{J.}~\bibnamefont{Linder}},
  \bibinfo{author}{\bibfnamefont{Y.}~\bibnamefont{Tanaka}},
  \bibinfo{author}{\bibfnamefont{T.}~\bibnamefont{Yokoyama}},
  \bibinfo{author}{\bibfnamefont{A.}~\bibnamefont{Sudb\o{}}}, \bibnamefont{and}
  \bibinfo{author}{\bibfnamefont{N.}~\bibnamefont{Nagaosa}},
  \bibinfo{journal}{Phys.\ Rev.\ Lett.} \textbf{\bibinfo{volume}{104}},
  \bibinfo{pages}{067001} (\bibinfo{year}{2010}).

\bibitem[{\citenamefont{Alicea}(2010)}]{Alicea}
\bibinfo{author}{\bibfnamefont{J.}~\bibnamefont{Alicea}},
  \bibinfo{journal}{Phys.\ Rev.\ B} \textbf{\bibinfo{volume}{81}},
  \bibinfo{pages}{125318} (\bibinfo{year}{2010}).

\bibitem[{\citenamefont{Mao et~al.}(2011)\citenamefont{Mao, Shi, Niu, and
  Zhang}}]{HoleDopedMajoranas}
\bibinfo{author}{\bibfnamefont{L.}~\bibnamefont{Mao}},
  \bibinfo{author}{\bibfnamefont{J.}~\bibnamefont{Shi}},
  \bibinfo{author}{\bibfnamefont{Q.}~\bibnamefont{Niu}}, \bibnamefont{and}
  \bibinfo{author}{\bibfnamefont{C.}~\bibnamefont{Zhang}},
  \bibinfo{journal}{Phys. Rev. Lett.} \textbf{\bibinfo{volume}{106}},
  \bibinfo{pages}{157003} (\bibinfo{year}{2011}).

\bibitem[{\citenamefont{Ghosh et~al.}(2010)\citenamefont{Ghosh, Sau, Tewari,
  and {Das Sarma}}}]{Ghosh}
\bibinfo{author}{\bibfnamefont{P.}~\bibnamefont{Ghosh}},
  \bibinfo{author}{\bibfnamefont{J.~D.} \bibnamefont{Sau}},
  \bibinfo{author}{\bibfnamefont{S.}~\bibnamefont{Tewari}}, \bibnamefont{and}
  \bibinfo{author}{\bibfnamefont{S.}~\bibnamefont{{Das Sarma}}},
  \bibinfo{journal}{Phys. Rev. B} \textbf{\bibinfo{volume}{82}},
  \bibinfo{pages}{184525} (\bibinfo{year}{2010}).

\bibitem[{\citenamefont{Qi et~al.}(2010)\citenamefont{Qi, Hughes, and
  Zhang}}]{Qi}
\bibinfo{author}{\bibfnamefont{X.-L.} \bibnamefont{Qi}},
  \bibinfo{author}{\bibfnamefont{T.~L.} \bibnamefont{Hughes}},
  \bibnamefont{and} \bibinfo{author}{\bibfnamefont{S.-C.} \bibnamefont{Zhang}},
  \bibinfo{journal}{Phys. Rev. B} \textbf{\bibinfo{volume}{82}},
  \bibinfo{pages}{184516} (\bibinfo{year}{2010}).

\bibitem[{\citenamefont{Lutchyn et~al.}(2010)\citenamefont{Lutchyn, Sau, and
  Das~Sarma}}]{1DwiresLutchyn}
\bibinfo{author}{\bibfnamefont{R.~M.} \bibnamefont{Lutchyn}},
  \bibinfo{author}{\bibfnamefont{J.~D.} \bibnamefont{Sau}}, \bibnamefont{and}
  \bibinfo{author}{\bibfnamefont{S.}~\bibnamefont{Das~Sarma}},
  \bibinfo{journal}{Phys.\ Rev.\ Lett.} \textbf{\bibinfo{volume}{105}},
  \bibinfo{pages}{077001} (\bibinfo{year}{2010}).

\bibitem[{\citenamefont{Oreg et~al.}(2010)\citenamefont{Oreg, Refael, and von
  Oppen}}]{1DwiresOreg}
\bibinfo{author}{\bibfnamefont{Y.}~\bibnamefont{Oreg}},
  \bibinfo{author}{\bibfnamefont{G.}~\bibnamefont{Refael}}, \bibnamefont{and}
  \bibinfo{author}{\bibfnamefont{F.}~\bibnamefont{von Oppen}},
  \bibinfo{journal}{Phys.\ Rev.\ Lett.} \textbf{\bibinfo{volume}{105}},
  \bibinfo{pages}{177002} (\bibinfo{year}{2010}).

\bibitem[{\citenamefont{Jiang et~al.}(unpublished)\citenamefont{Jiang,
  Kitagawa, Alicea, Akhmerov, Pekker, Refael, Cirac, Demler, Lukin, and
  Zoller}}]{ColdAtomMajoranas}
\bibinfo{author}{\bibfnamefont{L.}~\bibnamefont{Jiang}},
  \bibinfo{author}{\bibfnamefont{T.}~\bibnamefont{Kitagawa}},
  \bibinfo{author}{\bibfnamefont{J.}~\bibnamefont{Alicea}},
  \bibinfo{author}{\bibfnamefont{A.~R.} \bibnamefont{Akhmerov}},
  \bibinfo{author}{\bibfnamefont{D.}~\bibnamefont{Pekker}},
  \bibinfo{author}{\bibfnamefont{G.}~\bibnamefont{Refael}},
  \bibinfo{author}{\bibfnamefont{J.~I.} \bibnamefont{Cirac}},
  \bibinfo{author}{\bibfnamefont{E.}~\bibnamefont{Demler}},
  \bibinfo{author}{\bibfnamefont{M.~D.} \bibnamefont{Lukin}}, \bibnamefont{and}
  \bibinfo{author}{\bibfnamefont{P.}~\bibnamefont{Zoller}},
  \bibinfo{journal}{arXiv:1102.5367}  (\bibinfo{year}{unpublished}).

\bibitem[{\citenamefont{Weng et~al.}(unpublished)\citenamefont{Weng, Xu, Zhang,
  Zhang, Dai, and Fang}}]{HalfMetalNaCoO}
\bibinfo{author}{\bibfnamefont{H.}~\bibnamefont{Weng}},
  \bibinfo{author}{\bibfnamefont{G.}~\bibnamefont{Xu}},
  \bibinfo{author}{\bibfnamefont{H.}~\bibnamefont{Zhang}},
  \bibinfo{author}{\bibfnamefont{S.-C.} \bibnamefont{Zhang}},
  \bibinfo{author}{\bibfnamefont{X.}~\bibnamefont{Dai}}, \bibnamefont{and}
  \bibinfo{author}{\bibfnamefont{Z.}~\bibnamefont{Fang}},
  \bibinfo{journal}{arXiv:1103.1930}  (\bibinfo{year}{unpublished}).

\bibitem[{\citenamefont{Murakawa et~al.}(2011)\citenamefont{Murakawa, Wada,
  Tamura, Wasai, Saitoh, Aoki, Nomura, Okuda, Nagato, Yamamoto
  et~al.}}]{Murakawa:2011}
\bibinfo{author}{\bibfnamefont{S.}~\bibnamefont{Murakawa}},
  \bibinfo{author}{\bibfnamefont{Y.}~\bibnamefont{Wada}},
  \bibinfo{author}{\bibfnamefont{Y.}~\bibnamefont{Tamura}},
  \bibinfo{author}{\bibfnamefont{M.}~\bibnamefont{Wasai}},
  \bibinfo{author}{\bibfnamefont{M.}~\bibnamefont{Saitoh}},
  \bibinfo{author}{\bibfnamefont{Y.}~\bibnamefont{Aoki}},
  \bibinfo{author}{\bibfnamefont{R.}~\bibnamefont{Nomura}},
  \bibinfo{author}{\bibfnamefont{Y.}~\bibnamefont{Okuda}},
  \bibinfo{author}{\bibfnamefont{Y.}~\bibnamefont{Nagato}},
  \bibinfo{author}{\bibfnamefont{M.}~\bibnamefont{Yamamoto}},
  \bibnamefont{et~al.}, \bibinfo{journal}{Journal of the Physical Society of
  Japan} \textbf{\bibinfo{volume}{80}}, \bibinfo{pages}{013602}
  (\bibinfo{year}{2011}).

\bibitem[{\citenamefont{Doh et~al.}(2005)\citenamefont{Doh, {van Dam}, Roest,
  Bakkers, Kouwenhoven, and {De Franceschi}}}]{nanowireExpt}
\bibinfo{author}{\bibfnamefont{Y.-J.} \bibnamefont{Doh}},
  \bibinfo{author}{\bibfnamefont{J.~A.} \bibnamefont{{van Dam}}},
  \bibinfo{author}{\bibfnamefont{A.~L.} \bibnamefont{Roest}},
  \bibinfo{author}{\bibfnamefont{E.~P. A.~M.} \bibnamefont{Bakkers}},
  \bibinfo{author}{\bibfnamefont{L.~P.} \bibnamefont{Kouwenhoven}},
  \bibnamefont{and} \bibinfo{author}{\bibfnamefont{S.}~\bibnamefont{{De
  Franceschi}}}, \bibinfo{journal}{Science} \textbf{\bibinfo{volume}{309}},
  \bibinfo{pages}{272} (\bibinfo{year}{2005}).

\bibitem[{\citenamefont{Ivanov}(2001)}]{Ivanov}
\bibinfo{author}{\bibfnamefont{D.~A.} \bibnamefont{Ivanov}},
  \bibinfo{journal}{Phys.\ Rev.\ Lett.} \textbf{\bibinfo{volume}{86}},
  \bibinfo{pages}{268} (\bibinfo{year}{2001}).

\bibitem[{\citenamefont{Alicea et~al.}(2011)\citenamefont{Alicea, Oreg, Refael,
  {von Oppen}, and Fisher}}]{AliceaBraiding}
\bibinfo{author}{\bibfnamefont{J.}~\bibnamefont{Alicea}},
  \bibinfo{author}{\bibfnamefont{Y.}~\bibnamefont{Oreg}},
  \bibinfo{author}{\bibfnamefont{G.}~\bibnamefont{Refael}},
  \bibinfo{author}{\bibfnamefont{F.}~\bibnamefont{{von Oppen}}},
  \bibnamefont{and} \bibinfo{author}{\bibfnamefont{M.~P.~A.}
  \bibnamefont{Fisher}}, \bibinfo{journal}{Nature Physics, in press}
  (\bibinfo{year}{2011}).

\bibitem[{\citenamefont{Hassler et~al.}(2010)\citenamefont{Hassler, Akhmerov,
  Hou, and Beenakker}}]{Hassler}
\bibinfo{author}{\bibfnamefont{F.}~\bibnamefont{Hassler}},
  \bibinfo{author}{\bibfnamefont{A.~R.} \bibnamefont{Akhmerov}},
  \bibinfo{author}{\bibfnamefont{C.-Y.} \bibnamefont{Hou}}, \bibnamefont{and}
  \bibinfo{author}{\bibfnamefont{C.~W.~J.} \bibnamefont{Beenakker}},
  \bibinfo{journal}{New Journal of Physics} \textbf{\bibinfo{volume}{12}},
  \bibinfo{pages}{125002} (\bibinfo{year}{2010}).

\bibitem[{\citenamefont{Sau et~al.}(2010{\natexlab{b}})\citenamefont{Sau,
  Tewari, and {Das Sarma}}}]{SauWireNetwork}
\bibinfo{author}{\bibfnamefont{J.~D.} \bibnamefont{Sau}},
  \bibinfo{author}{\bibfnamefont{S.}~\bibnamefont{Tewari}}, \bibnamefont{and}
  \bibinfo{author}{\bibfnamefont{S.}~\bibnamefont{{Das Sarma}}},
  \bibinfo{journal}{Phys. Rev. A} \textbf{\bibinfo{volume}{82}},
  \bibinfo{pages}{052322} (\bibinfo{year}{2010}{\natexlab{b}}).

\bibitem[{\citenamefont{Bonderson and Lutchyn}(2011)}]{TopologicalQuantumBus}
\bibinfo{author}{\bibfnamefont{P.}~\bibnamefont{Bonderson}} \bibnamefont{and}
  \bibinfo{author}{\bibfnamefont{R.~M.} \bibnamefont{Lutchyn}},
  \bibinfo{journal}{Phys. Rev. Lett.} \textbf{\bibinfo{volume}{106}},
  \bibinfo{pages}{130505} (\bibinfo{year}{2011}).

\bibitem[{\citenamefont{Flensberg}(2011)}]{Flensberg}
\bibinfo{author}{\bibfnamefont{K.}~\bibnamefont{Flensberg}},
  \bibinfo{journal}{Phys. Rev. Lett.} \textbf{\bibinfo{volume}{106}},
  \bibinfo{pages}{090503} (\bibinfo{year}{2011}).

\bibitem[{\citenamefont{Potter and Lee}(2010)}]{Multichannel1}
\bibinfo{author}{\bibfnamefont{A.~C.} \bibnamefont{Potter}} \bibnamefont{and}
  \bibinfo{author}{\bibfnamefont{P.~A.} \bibnamefont{Lee}},
  \bibinfo{journal}{Phys. Rev. Lett.} \textbf{\bibinfo{volume}{105}},
  \bibinfo{pages}{227003} (\bibinfo{year}{2010}).

\bibitem[{\citenamefont{Lutchyn et~al.}(2011)\citenamefont{Lutchyn, Stanescu,
  and {Das Sarma}}}]{Multichannel2}
\bibinfo{author}{\bibfnamefont{R.~M.} \bibnamefont{Lutchyn}},
  \bibinfo{author}{\bibfnamefont{T.~D.} \bibnamefont{Stanescu}},
  \bibnamefont{and} \bibinfo{author}{\bibfnamefont{S.}~\bibnamefont{{Das
  Sarma}}}, \bibinfo{journal}{Phys. Rev. Lett.} \textbf{\bibinfo{volume}{106}},
  \bibinfo{pages}{127001} (\bibinfo{year}{2011}).

\bibitem[{\citenamefont{Potter and Lee}(2011)}]{Multichannel3}
\bibinfo{author}{\bibfnamefont{A.~C.} \bibnamefont{Potter}} \bibnamefont{and}
  \bibinfo{author}{\bibfnamefont{P.~A.} \bibnamefont{Lee}},
  \bibinfo{journal}{Phys. Rev. B} \textbf{\bibinfo{volume}{83}},
  \bibinfo{pages}{094525} (\bibinfo{year}{2011}).

\bibitem[{\citenamefont{Flensberg}(2010)}]{Disorder1}
\bibinfo{author}{\bibfnamefont{K.}~\bibnamefont{Flensberg}},
  \bibinfo{journal}{Phys. Rev. B} \textbf{\bibinfo{volume}{82}},
  \bibinfo{pages}{180516} (\bibinfo{year}{2010}).

\bibitem[{\citenamefont{Akhmerov et~al.}(2011)\citenamefont{Akhmerov, Dahlhaus,
  Hassler, Wimmer, and Beenakker}}]{Disorder2}
\bibinfo{author}{\bibfnamefont{A.~R.} \bibnamefont{Akhmerov}},
  \bibinfo{author}{\bibfnamefont{J.~P.} \bibnamefont{Dahlhaus}},
  \bibinfo{author}{\bibfnamefont{F.}~\bibnamefont{Hassler}},
  \bibinfo{author}{\bibfnamefont{M.}~\bibnamefont{Wimmer}}, \bibnamefont{and}
  \bibinfo{author}{\bibfnamefont{C.~W.~J.} \bibnamefont{Beenakker}},
  \bibinfo{journal}{Phys. Rev. Lett.} \textbf{\bibinfo{volume}{106}},
  \bibinfo{pages}{057001} (\bibinfo{year}{2011}).

\bibitem[{\citenamefont{Fulga et~al.}(unpublished)\citenamefont{Fulga, Hassler,
  Akhmerov, and Beenakker}}]{Disorder3}
\bibinfo{author}{\bibfnamefont{I.~C.} \bibnamefont{Fulga}},
  \bibinfo{author}{\bibfnamefont{F.}~\bibnamefont{Hassler}},
  \bibinfo{author}{\bibfnamefont{A.~R.} \bibnamefont{Akhmerov}},
  \bibnamefont{and} \bibinfo{author}{\bibfnamefont{C.~W.~J.}
  \bibnamefont{Beenakker}}, \bibinfo{journal}{arXiv:1101.1749}
  (\bibinfo{year}{unpublished}).

\bibitem[{\citenamefont{Potter and Lee}(unpublished)}]{Disorder4}
\bibinfo{author}{\bibfnamefont{A.~C.} \bibnamefont{Potter}} \bibnamefont{and}
  \bibinfo{author}{\bibfnamefont{P.~A.} \bibnamefont{Lee}},
  \bibinfo{journal}{arXiv:1103.2129}  (\bibinfo{year}{unpublished}).

\bibitem[{\citenamefont{Brouwer
  et~al.}(unpublished{\natexlab{a}})\citenamefont{Brouwer, Duckheim, Romito,
  and {von Oppen}}}]{Disorder5}
\bibinfo{author}{\bibfnamefont{P.~W.} \bibnamefont{Brouwer}},
  \bibinfo{author}{\bibfnamefont{M.}~\bibnamefont{Duckheim}},
  \bibinfo{author}{\bibfnamefont{A.}~\bibnamefont{Romito}}, \bibnamefont{and}
  \bibinfo{author}{\bibfnamefont{F.}~\bibnamefont{{von Oppen}}},
  \bibinfo{journal}{arXiv:1103.2746}
  (\bibinfo{year}{unpublished}{\natexlab{a}}).

\bibitem[{\citenamefont{Brouwer
  et~al.}(unpublished{\natexlab{b}})\citenamefont{Brouwer, Duckheim, Romito,
  and {von Oppen}}}]{Disorder6}
\bibinfo{author}{\bibfnamefont{P.~W.} \bibnamefont{Brouwer}},
  \bibinfo{author}{\bibfnamefont{M.}~\bibnamefont{Duckheim}},
  \bibinfo{author}{\bibfnamefont{A.}~\bibnamefont{Romito}}, \bibnamefont{and}
  \bibinfo{author}{\bibfnamefont{F.}~\bibnamefont{{von Oppen}}},
  \bibinfo{journal}{arXiv:1104.1531}
  (\bibinfo{year}{unpublished}{\natexlab{b}}).

\bibitem[{\citenamefont{Sun et~al.}(2007)\citenamefont{Sun, Gangadharaiah, and
  Starykh}}]{Sun:2007}
\bibinfo{author}{\bibfnamefont{J.}~\bibnamefont{Sun}},
  \bibinfo{author}{\bibfnamefont{S.}~\bibnamefont{Gangadharaiah}},
  \bibnamefont{and} \bibinfo{author}{\bibfnamefont{O.~A.}
  \bibnamefont{Starykh}}, \bibinfo{journal}{Phys. Rev. Lett.}
  \textbf{\bibinfo{volume}{98}}, \bibinfo{pages}{126408}
  (\bibinfo{year}{2007}).

\bibitem[{\citenamefont{Gangadharaiah et~al.}(2008)\citenamefont{Gangadharaiah,
  Sun, and Starykh}}]{Gangadharaiah:2008}
\bibinfo{author}{\bibfnamefont{S.}~\bibnamefont{Gangadharaiah}},
  \bibinfo{author}{\bibfnamefont{J.}~\bibnamefont{Sun}}, \bibnamefont{and}
  \bibinfo{author}{\bibfnamefont{O.~A.} \bibnamefont{Starykh}},
  \bibinfo{journal}{Phys. Rev. B} \textbf{\bibinfo{volume}{78}},
  \bibinfo{pages}{054436} (\bibinfo{year}{2008}).

\bibitem[{\citenamefont{Gangadharaiah
  et~al.}(unpublished)\citenamefont{Gangadharaiah, Braunecker, Simon, and
  Loss}}]{Gangadharaiah:2011}
\bibinfo{author}{\bibfnamefont{S.}~\bibnamefont{Gangadharaiah}},
  \bibinfo{author}{\bibfnamefont{B.}~\bibnamefont{Braunecker}},
  \bibinfo{author}{\bibfnamefont{P.}~\bibnamefont{Simon}}, \bibnamefont{and}
  \bibinfo{author}{\bibfnamefont{D.}~\bibnamefont{Loss}},
  \bibinfo{journal}{arXiv:1101.0094}  (\bibinfo{year}{unpublished}).

\bibitem[{\citenamefont{Sela et~al.}(unpublished)\citenamefont{Sela, Altland,
  and Rosch}}]{Sela:2011}
\bibinfo{author}{\bibfnamefont{E.}~\bibnamefont{Sela}},
  \bibinfo{author}{\bibfnamefont{A.}~\bibnamefont{Altland}}, \bibnamefont{and}
  \bibinfo{author}{\bibfnamefont{A.}~\bibnamefont{Rosch}},
  \bibinfo{journal}{arXiv:1103.4969}  (\bibinfo{year}{unpublished}).

\bibitem[{\citenamefont{Kitaev}(2001)}]{Kitaev:2001}
\bibinfo{author}{\bibfnamefont{A.~Y.} \bibnamefont{Kitaev}},
  \bibinfo{journal}{Physics-Uspekhi} \textbf{\bibinfo{volume}{44}},
  \bibinfo{pages}{131} (\bibinfo{year}{2001}).

\bibitem[{\citenamefont{Turner et~al.}(unpublished)\citenamefont{Turner,
  Pollmann, and Berg}}]{Turner:2010}
\bibinfo{author}{\bibfnamefont{A.~M.} \bibnamefont{Turner}},
  \bibinfo{author}{\bibfnamefont{F.}~\bibnamefont{Pollmann}}, \bibnamefont{and}
  \bibinfo{author}{\bibfnamefont{E.}~\bibnamefont{Berg}},
  \bibinfo{journal}{arXiv:1008.4346}  (\bibinfo{year}{unpublished}).

\bibitem[{\citenamefont{Qi et~al.}(unpublished)\citenamefont{Qi, Katsura, and
  Ludwig}}]{Qi:2011}
\bibinfo{author}{\bibfnamefont{X.-L.} \bibnamefont{Qi}},
  \bibinfo{author}{\bibfnamefont{H.}~\bibnamefont{Katsura}}, \bibnamefont{and}
  \bibinfo{author}{\bibfnamefont{A.}~\bibnamefont{Ludwig}},
  \bibinfo{journal}{arXiv:1103.5437}  (\bibinfo{year}{unpublished}).

\bibitem[{\citenamefont{Schollw\"ock}(2011)}]{Schollwoeck:2011}
\bibinfo{author}{\bibfnamefont{U.}~\bibnamefont{Schollw\"ock}},
  \bibinfo{journal}{Annals of Physics} \textbf{\bibinfo{volume}{326}},
  \bibinfo{pages}{96} (\bibinfo{year}{2011}).

\bibitem[{\citenamefont{Giamarchi}(2004)}]{Giamarchi:2004}
\bibinfo{author}{\bibfnamefont{T.}~\bibnamefont{Giamarchi}},
  \emph{\bibinfo{title}{Quantum Physics in One Dimension}}
  (\bibinfo{publisher}{Oxford University Press}, \bibinfo{year}{2004}).

\bibitem[{\citenamefont{Shekhtman et~al.}(1992)\citenamefont{Shekhtman,
  Entin-Wohlman, and Aharony}}]{Shekhtman:1992}
\bibinfo{author}{\bibfnamefont{L.}~\bibnamefont{Shekhtman}},
  \bibinfo{author}{\bibfnamefont{O.}~\bibnamefont{Entin-Wohlman}},
  \bibnamefont{and} \bibinfo{author}{\bibfnamefont{A.}~\bibnamefont{Aharony}},
  \bibinfo{journal}{Phys. Rev. Lett.} \textbf{\bibinfo{volume}{69}},
  \bibinfo{pages}{836} (\bibinfo{year}{1992}).

\bibitem[{\citenamefont{Schulz}(1990)}]{LuttingerParameterHubbardModel}
\bibinfo{author}{\bibfnamefont{H.~J.} \bibnamefont{Schulz}},
  \bibinfo{journal}{Phys. Rev. Lett.} \textbf{\bibinfo{volume}{64}},
  \bibinfo{pages}{2831} (\bibinfo{year}{1990}).

\bibitem[{\citenamefont{Starykh et~al.}(2005)\citenamefont{Starykh, Furusaki,
  and Balents}}]{starykh05}
\bibinfo{author}{\bibfnamefont{O.~A.} \bibnamefont{Starykh}},
  \bibinfo{author}{\bibfnamefont{A.}~\bibnamefont{Furusaki}}, \bibnamefont{and}
  \bibinfo{author}{\bibfnamefont{L.}~\bibnamefont{Balents}},
  \bibinfo{journal}{Phys. Rev. B} \textbf{\bibinfo{volume}{72}},
  \bibinfo{pages}{094416} (\bibinfo{year}{2005}).

\bibitem[{\citenamefont{Lieb and Wu}(1968)}]{Lieb:1968}
\bibinfo{author}{\bibfnamefont{E.}~\bibnamefont{Lieb}} \bibnamefont{and}
  \bibinfo{author}{\bibfnamefont{F.}~\bibnamefont{Wu}}, \bibinfo{journal}{Phys.
  Rev. Lett.} \textbf{\bibinfo{volume}{20}}, \bibinfo{pages}{1445}
  (\bibinfo{year}{1968}).

\bibitem[{\citenamefont{Wimmer et~al.}(2010)\citenamefont{Wimmer, Akhmerov,
  Medvedyeva, Tworzyd\l{}o, and Beenakker}}]{WimmerMultichannel}
\bibinfo{author}{\bibfnamefont{M.}~\bibnamefont{Wimmer}},
  \bibinfo{author}{\bibfnamefont{A.~R.} \bibnamefont{Akhmerov}},
  \bibinfo{author}{\bibfnamefont{M.~V.} \bibnamefont{Medvedyeva}},
  \bibinfo{author}{\bibfnamefont{J.}~\bibnamefont{Tworzyd\l{}o}},
  \bibnamefont{and} \bibinfo{author}{\bibfnamefont{C.~W.~J.}
  \bibnamefont{Beenakker}}, \bibinfo{journal}{Phys. Rev. Lett.}
  \textbf{\bibinfo{volume}{105}}, \bibinfo{pages}{046803}
  (\bibinfo{year}{2010}).

\bibitem[{\citenamefont{Law and Lee}(unpublished)}]{Multichannel4}
\bibinfo{author}{\bibfnamefont{K.~T.} \bibnamefont{Law}} \bibnamefont{and}
  \bibinfo{author}{\bibfnamefont{P.~A.} \bibnamefont{Lee}},
  \bibinfo{journal}{arXiv:1103.5013}  (\bibinfo{year}{unpublished}).

\bibitem[{\citenamefont{Lutchyn and Fisher}(unpublished)}]{Lutchyn:2011}
\bibinfo{author}{\bibfnamefont{R.~M.} \bibnamefont{Lutchyn}} \bibnamefont{and}
  \bibinfo{author}{\bibfnamefont{M.~P.~A.} \bibnamefont{Fisher}},
  \bibinfo{journal}{arXiv:1104.2358}  (\bibinfo{year}{unpublished}).

\end{thebibliography}

\end{document}